\documentclass[preprint,12pt,3p]{elsarticle}
\journal{Climate Action}

\usepackage{matlab-prettifier}
\usepackage{graphicx}
\usepackage{amsmath}
\usepackage{xfrac}
\usepackage{parskip}
\usepackage[backref=page]{hyperref}
\usepackage{adjustbox}
\usepackage{setspace}
\hypersetup{ 
backref=true, 
bookmarks=true, 
unicode=false, 
pdftoolbar=true, 
pdfmenubar=true, 
pdffitwindow=false, 
pdfstartview={FitH}, 
pdftitle={}, 
pdfauthor={}, 
pdfsubject={}, 
pdfkeywords={}{}, 
pdfnewwindow=true, 
colorlinks=true, 
linkcolor=BlueViolet, 
citecolor=maroon, 
urlcolor=blue,
}
\usepackage{comment}
\usepackage{subfig}
\usepackage{caption}
\usepackage{longtable}
\usepackage{amssymb}
\usepackage{gensymb}
\usepackage{siunitx}

\usepackage{natbib}
\biboptions{comma,round}
\setcitestyle{authoryear}

\begin{document}
\begin{frontmatter}

\title{The IPCC and the challenge of \textit{ex post} policy evaluation}

\author[label1,label2,label3,label4,label5,label6]{Richard S.J. Tol\corref{cor1}\fnref{label8}}
\address[label1]{Department of Economics, University of Sussex, Falmer, United Kingdom}
\address[label2]{Institute for Environmental Studies, Vrije Universiteit, Amsterdam, The Netherlands}
\address[label3]{Department of Spatial Economics, Vrije Universiteit, Amsterdam, The Netherlands}
\address[label4]{Tinbergen Institute, Amsterdam, The Netherlands}
\address[label5]{CESifo, Munich, Germany}
\address[label6]{Payne Institute for Public Policy, Colorado School of Mines, Golden, CO, USA}

\cortext[cor1]{Jubilee Building, BN1 9SL, UK}
\fntext[labe8]{Elin Lerum Boasson, Erlend Hermansen, Glen Peters, and Ida Sognn\ae s had excellent comments on a previous version. Unreferenced factual statements are first-hand observations by the author.}

\ead{r.tol@sussex.ac.uk}
\ead[url]{http://www.ae-info.org/ae/Member/Tol\_Richard}

\begin{abstract}
The IPCC started at a time when climate policy was an aspiration for the future. The research assessed in the early IPCC reports was necessarily about \emph{potential} climate policies, always stylized and often optimized. The IPCC has continued on this path, even though there is now a considerable literature studying \emph{actual} climate policy, in all its infuriating detail, warts and all. Four case studies suggest that the IPCC, in its current form, will not be able to successfully switch from \textit{ex ante} to \textit{ex post} policy evaluation. This transition is key as AR7 will most likely have to confront the failure to meet the 1.5\celsius{} target. The four cases are as follows. (1) The scenarios first build and later endorsed by the IPCC all project a peaceful future with steady if not rapid economic growth everywhere, more closely resembling political manifestos than facts on the ground. (2) Successive IPCC reports have studiously avoided discussing the voluminous literature suggesting that political targets for greenhouse gas emission reduction are far from optimal, although a central part of that work was awarded the Nobel Prize in 2018. (3) IPCC AR5 found it impossible to acknowledge that the international climate policy negotiations from COP1 (Berlin) to COP19 (Warsaw) were bound to fail, just months before the radical overhaul at COP20 (Lima) proved that point. (4) IPCC AR6 by and large omitted the nascent literature on \textit{ex post} climate policy evaluation. Together, these cases suggest that the IPCC finds self-criticism difficult and is too close to policy makers to criticize past and current policy mistakes. One solution would be to move control over the IPCC to the national authorities on research and higher education.
\end{abstract}

\end{frontmatter}

\section{Introduction}
Future reports of the Intergovernmental Panel on Climate Change (IPCC) will have to assess the successes and failures of climate policy and the IPCC's predictions about climate policy. I argue that, as currently constituted, the IPCC will find it difficult to do so.

The IPCC was founded in 1988 ``to provide governments at all levels with scientific information that they can use to develop climate policies.'' At the time, there was no climate policy to speak of. The academic literature could only consider potential future policies. The First Assessment Report (AR1) of the IPCC, published in 1990, was necessarily limited to \textit{ex ante} policy evaluations\textemdash the first carbon taxes were levied in Poland and Finland in 1990.

The Sixth Assessment Report (AR6) was still by and large focused on future climate policies, even though there is now over 30 years of experience with greenhouse gas emission reduction and a rich and growing literature evaluating the successes, or lack thereof, of these policies. There is also a nascent literature comparing the \emph{forecast} impact of climate policy to its \emph{actual} impact.

The ``I'' in IPCC is for \emph{intergovernmental}. IPCC reports are signed off by the \emph{Panel}, which consist of government representatives. Government officials tend to grudgingly accept warnings by the IPCC that current and planned policy interventions are insufficiently ambitious or suboptimally designed. Planned policy is always in flux and spin doctors work hard to make things seem better than they are. Governments do not much mind criticism of their plans. However, officials are less keen to announce to the world that past policies did not meet their stated goals, made things worse, or were corrupt.

In the Seventh Assessment Report (AR7), the IPCC will have to come to terms with the learned papers that study past climate policy. That literature is growing rapidly (see below) and will soon be too large to ignore. It is also likely that, by the end of the decade when AR7 is expected,\footnote{The publication date for AR7 has yet to be announced. The period between assessment reports has grown longer over time.} the 1.5\celsius{} warming target of the Paris Agreement will have been breached \citep{WMO2022} and emissions will not have fallen nearly as much as promised \citep{UNFCCC2021}. AR7 simply cannot avoid discussing policy failure.

As part of that, the IPCC will also need to discuss those models that forecast that climate policy would be easy and cheap. Some of these models were championed in previous IPCC reports.

The IPCC will also have to confront those who predicted catastrophe or worse if the world warms by more than 1.5\celsius.\footnote{The literature on failed prophecies suggests that the prophets of doom will not lose credibility in the eyes of their acolytes \citep{Festinger1956, Melton1985, Dawson1999, Dein2001}.} This is a somewhat easier task as the IPCC has suggested doom and gloom between the lines, but never outright predicted the apocalypse. The IPCC, however, did make forecasts for what would happen at 1.5\celsius{} warming, forecasts that can be checked against data before the end of the decade.

Is the IPCC equal to these tasks? Much has been written about the IPCC \citep{Agrawala1998a, Agrawala1998b, Edwards2001, Alexander2007, Nishioka2008, Rothman2009, InterAcademy2010, Hulme2010, Nature2010, Carraro2015, Chan2016}, but not about the issues addressed in this paper. I here focus on the question whether the IPCC can successfully transition from \textit{ex ante} to \textit{ex post} policy evaluation. This is necessarily speculative. However, there are episodes in the IPCC's past that suggest it will not be able to cope, not with \textit{ex post} studies per se, but rather with the implications for the IPCC itself and its relations with policy makers.

In the rest of this paper, I first look, in Section \ref{sc:cases}, at four cases which together show that the IPCC, as currently conceived, is not capable of speaking truth to power. The IPCC can urge governments to adjust future plans, but it has found it difficult to criticize its paymasters for their past and current actions. The case studies also demonstrate that the IPCC finds it just as hard to critique its own work. I then discuss, in Section \ref{sc:remedies}, how the structure of the IPCC can be changed to facilitate \textit{ex post} policy evaluation before drawing conclusions in Section \ref{sc:conclude}.

\section{Four cases}
\label{sc:cases}

\subsection{Scenarios}
Scenarios are not implausible, internally consistent descriptions of alternative futures. The IPCC used to construct its own scenarios of population, economic activity, and greenhouse gas emissions \citep[][see also \citet{ONeill2008a, ONeill2008b, Girod2009}]{Tirpak1990, Leggett1992, SRES}. However, after these scenarios were criticised for the way exchange rates were handled, scenarios have been developed at arms' length of the IPCC \citep{Moss2010, vanVuuren2011, ONeill2014, Riahi2017}.\footnote{These arms are not very long. Many of the people who hold key positions in scenario development also hold key positions in the IPCC \citep{Cointe2019}. These \textit{de jure} non-IPCC but \textit{de facto} IPCC scenarios have the legitimacy of the IPCC without its constraints. This is deliberate.} The current set of scenarios, Representative Concentration Pathways (RCPs) and Shared Socio-economic Pathways (SSPs) is not uncontroversial either: The high concentrations scenarios make assumptions about the energy market that are at odds with technological progress and relative price trends over the last decade.

The earlier controversy over IPCC scenarios was started by two well-respected statisticians, who noted that international comparisons of standards of living and projections of future income were done using market-exchange dollars rather than more appropriate international or Geary-Khamis dollars \citep{Castles2003a, Castles2003b}. The difference arises because some goods and services are supplied locally or nationally, rather than on the world market, and the purchasing power of local currency is often higher than it would appear on first sight\textemdash as any rich-world traveller to a poorer country can attest.

The initial critique was overstated. Emission accounts are based on \emph{physical} accounts of the volume of energy use, not on \emph{economic} accounts of its value. A switch from market to international dollars would thus be offset by a recalibration of model parameters, notably income elasticities \citep{Holtsmark2004, Holtsmark2005, Manne2005, Tol2006, VanVuuren2006}.

However, this episode revealed two things. First, IPCC scenarios and the big integrated assessment models used to quantify these scenarios are build by people who are not economists, even though economic growth is a key driver of future emissions. This is manifest in the accounting error that was uncovered by Castles and Henderson but could have been spotted by an undergraduate student. Furthermore, the difference between market and international dollars is so profound because IPCC scenarios assume that incomes across the world will eventually converge. This assumption of $\sigma$-convergence has long been abandoned by economists\textemdash the data show otherwise \citep{Barro1992}\textemdash but remains a feature of the IPCC. Again, this is something taught to undergraduate economists. Another aspect of these scenarios that would surprise undergraduate students is that population growth is assumed to be independent of economic growth. Anyone would be amazed to learn that there are no wars, civil or international, in these scenarios.

The second revelation is the response of the IPCC as a body and of key people active in the organization. Instead of engaging with the critique, the initial response was to minimize the issue, to ridicule the critique, and to attack the critics \citep{Nakicenovic2003, Grubler2004}. Only after considerable push-back that Castles and Henderson did have a point, albeit not quite the point that they thought they had, did the IPCC boffins admit that their exchange rate accounting could be improved\textemdash in the \href{https://data.ece.iiasa.ac.at/ar6/}{IPCC AR6 scenario database}, some models no longer report income using market exchange rates. The root causes were never addressed, however: Few economists are involved in constructing the scenarios of the economy used by the IPCC.\footnote{In a companion paper, Ilan Noy calls for more economists to a be involved in the IPCC. This discussion and the ones below would support this, although it would be wrong for the IPCC to be dominated by any one discipline. Indeed, in another companion paper, Terry Barker Hector Pollitt lament the influence of orthodox economists on the IPCC, although it is hard to see how an organization like the IPCC could ever embrace heterodox schools of thought.}

Whereas the first three sets of scenarios were developed directly by the IPCC, the fourth and current set of scenarios were developed at one remove. The IPCC was never supposed to do primary research\textemdash scenario development is research. Creating distance between the IPCC and the scenarios it uses deflects criticism. The scenario builders and the IPCC authors assessing the scenarios are drawn from the same group of experts, however, if not the same people.

The current scenarios, SSPs, retain three features of the previous ones: Populations grow independent of economic circumstances, incomes converge, and world peace breaks out.\footnote{IPCC WG3 AR6 uses a wider range of scenarios than the SSPs, but these appear to have much the same features.} The first feature is explained by a failure to break-down the walls between disciplines. Instead of integrating demography into integrated assessment models, population models stand alone. Rapid economic growth in poor countries is assumed because developing countries are the majority in the IPCC panel. Accepting a report that projects economic stagnation in parts of the world could be interpreted as accepting policy failure by the responsible governments. For the same reason, the IPCC would not dare to forecast war or civil war, even if its impact on emissions could be stark, as illustrated by \citet{Tol2013China} for China and as is currently observed for the second Russian invasion of Ukraine \citep{Pereira2022}.

Recent critiques focus on a different aspect of the scenarios \citep{Pielke2008, Richels2008, Ritchie2017, Burgess2020, Pielke2021, Pielke2022, Srikrishnan2022, Yuan2022}. Reserves and resources of conventional oil and gas are likely to be exhausted in the 21st century and will need to be replaced by another energy source. This is a key part of any emissions scenario: If oil and gas are replaced by renewables and nuclear, emissions would fall; if unconventional oil and gas is the future fuel of choice, emissions would rise slightly; if coal returns as the mainstay of the energy sector, perhaps liquefied or gasified, emissions would rise rapidly.

The high emission scenarios in the SSP set assume a return to coal. This may have been somewhat credible when these scenarios were build. Since then, developments in shale oil and gas, in wind and solar power, in batteries, and in the electrification of transport and heating make a resurgence of coal hard to imagine. Scenarios are supposed to be not-implausible; coal no longer meets that bar. This is underlined by the amounts of coal used in these scenarios, which may require mining coal in Antarctica \citep{Ritchie2017}\textemdash which is not just forbidden by international treaty but also prohibitively expensive.

Valid criticism notwithstanding, the high SRES scenarios dominate IPCC AR6 with scary a word of reflection. Indeed, these scenarios are all too often described as ``business as usual'' or ``no (additional) policy'' scenarios, which is plainly wrong. Aware of the critique, the IPCC decided to stay mostly silent rather than appropriately caveating their use of scenarios and calling for a reappraisal.\footnote{It takes a number of years from scenario development to its deployment in studies of climate change, climate policy, and its impacts. Rapid rebalancing is not an option for the IPCC. Unitary organizations with shorter cause-effect chains, such as the International Energy Agency, can update their scenarios on an annual basis.}

This case shows that the IPCC has found it difficult to admit past mistakes.

\subsection{Optimal climate policy}
The first cost-benefit analysis of climate policy was published 6 years before the IPCC was formed \citep{Nordhaus1982}. It was largely ignored. The second study of optimal emission reduction was published shortly after AR1 \citep{Nordhaus1992}. This paper spawned a large and complex literature on a question that is at the heart of climate policy, namely how to balance the costs of greenhouse gas emission reduction against the benefits of avoided climate change, a literature that led to the first Nobel Prize being awarded to someone who had studied climate change for most of his career.\footnote{Klaus Hasselmann and Syukuro Manabe won three years after William Nordhaus. The 2007 Nobel Peace Prize was awarded to the IPCC.} The literature on cost-benefit analysis of climate policy includes many papers arguing against Nordhaus' key conclusions\textemdash indeed, trying to prove Nordhaus wrong has been a mission for three generations of climate economists. You would think that this intellectually exciting, policy-relevant research would take pride of place in the assessment reports of the IPCC. You would be wrong.

Economists were largely absent from AR1. In AR2, however, they made their presence felt. The social cost chapter \citep{Pearce1996IPCC} attracted worldwide attention and controversy \citep{Bruce1995, Masood1995a, Masood1995b, Meyer1995a, Meyer1995b, Nature1995, Pearce1995c, Pearce1995a, Pearce1995b, Courtney1996, Grubb1996, Fankhauser1997b, Tol1997, Fankhauser1998}. Economists routinely value risks to human health based on people's willingness to pay to reduce such risks, and willingness to pay is constrained by ability to pay. Richer people thus appear to be worth more. The governments of poorer countries, India in particular, took affront. Instead of defending the academic literature against political objections, the IPCC Bureau sided with the IPCC Plenary. The Summary for Policy Makers disowns the chapter.\footnote{Having served in various roles for the US government, Nordhaus is not adverse to policy advice. Always cautious, he did not participate in AR2 but closely observed what went on. He declined to serve in AR3 and later reports. Two Nobelists, Arrow and Stiglitz, did serve in AR2 but were not thrilled by the political interference either. No economist of similar stature has been part of the IPCC after that. Economists did return to the IPCC for AR5, with similar results. See Section \ref{ssc:games}.} 

A second controversy played out away from the media. The government of Switzerland protested the inclusion of a chapter on integrated assessment modelling \citep{Weyant1996IPCC}, ostensibly on procedural grounds: The chapter was not part of the initial, approved outline. Switzerland had no problem with the chapter on discounting \citep{Arrow1996IPCC}, which too was added later. In private, Swiss officials admitted that their objection was not procedural. Instead, they argued that cost-benefit analysis with integrated assessment models gives \emph{direct} policy advice, telling governments what the carbon tax should be. Switzerland preferred policy advice over technocracy, a preference shared more widely. Most policy analysts would agree with the Swiss delegates: Cost-benefit analysis informs but not proscribes policy; economists are plumbers \citep{Duflo2017} not effective altruists. Cost-benefit analysis is one way to systematically work through the pros and cons of various policy options, and to highlight the trade-offs that are part of any decision on a complex issue.

Switzerland did not succeed in excluding the integrated assessment chapter from AR2. However, the restructuring of the IPCC after AR2 excludes this work by default. The IPCC is currently split into three working groups. Working Group I is on the natural science of climate change, Working Group II on the impacts of climate change, and Working Group III on climate policy. The Synthesis Report brings the three strands together, but cannot add new material. Cost-benefit analysis of climate policy fits in neither WG2 nor in WG3 and is therefore excluded from the IPCC.

AR3 and AR5 included a discussion of the economic impacts of climate change, including the social cost of carbon or recommended carbon tax \citep[][see also \citet{Tol2016}]{Smith2001, Arent2014} but this again ran into political objections. The SPM and underlying chapter diverge. The impacts of climate change as estimated by economists are (a lot) lower than many people think they are or should be. The IPCC seems unable or unwilling to correct widespread misperceptions. AR6 has a hastily prepared ``cross working group paper'' on the topic \citep{Rose2022} that too was omitted from the Summary for Policy Makers. The IPCC once again ignored politically inconvenient research.

The economics literature is inconvenient because optimal emission reduction is less stringent than political aspirations \citep{Tol1999kyoto, Tol2012EP, Tol2013JECD, Tol2021IE}. Never mind that the recommended carbon tax has been higher than the observed carbon price for 40 years \citep{Nordhaus1982, Tol2018REEP}. The structure of cost-benefit analysis implies that the stabilization of the atmospheric concentration of carbon dioxide cannot be optimal \citep{Tol2019bk}. Optimization models that stabilize the climate do this by virtue of a carbon-free backstop technology, introduced by \citet{Nordhaus2014}, rather than because of a concern about climate change. The impact of climate change sets only the \emph{level} at which the climate is stabilized \citep{Hansel2020}.

This case shows that the IPCC finds it difficult to publish politically inconvenient results.

\subsection{Game theory}
\label{ssc:games}
The same tendency to avoid politically inconvenient conclusions was on display in another part of AR5. Greenhouse gas emission reduction is a public good. If one country reduces its emissions, other countries enjoy the benefits of less climate change. That is, emission reduction is non-rival. No country can stop another country from cutting its emissions. That is, emission reduction is non-excludable. Non-excludability and non-rivalry together imply that greenhouse gas emission reduction is a public good. Public goods are underprovided by the free market, an insight that goes back to Adam \citet{Smith1776}. Governments need to provide public goods. Some even go as far as saying that government was invented for this purpose. Greenhouse gas emission reduction is a \emph{global} public good, but there is no \emph{world} government to provide it.\footnote{Political scientists approach the formation of international environmental agreements differently, but agree that environmentally meaningful treaties are few and far between.}

From the first Conference of the Parties (COP) to the United Nations Framework on Climate Change (UNFCCC) in Berlin in 1995 to COP19 in 2013 in Warsaw, countries tried to negotiate legally binding targets for emission reduction. As countries are sovereign, this is tantamount to the private provision of a public good. Just how difficult this is can be seen from the 1997 Kyoto Protocol\textemdash few countries signed up and some who did walked away later\textemdash and the acrimony in 2009 in Copenhagen.

The difficulty in forming an international agreement on greenhouse gas emission reduction was foreseen before countries started negotiating \citep{Carraro1992, Carraro1993, Barrett1994}. A steady stream of papers has since tested the robustness of this conclusion and by and large confirmed the earlier findings \citep[see][for a review and an exception]{Battaglini2016}. The penultimate draft of the Summary for Policy Makers of AR5 synthesized this large, complex, and technical literature with stark clarity and brevity.\footnote{It is no longer available. The author, Rob Stavins, was initially part of AR6 but withdrew. I have not asked him why.} It did not survive contact with government representatives \citep{Stavins2014, Stavins2015}, apparently loath to admit that 20 years of negotiations had been in vain\textemdash or perhaps they were taken aback by the implied ``I told you so''.

AR5 was concluded in 2014. In December of the same year, in Lima, the international climate negotiations ended the attempt to negotiate binding targets, replacing them by intended nationally determined contributions. This pledge-and-review system, favoured by economists \citep{Bradford2008}, is a key part of the 2015 Paris Agreement. The Lima Conversion had been years in the making and had been openly discussed by senior negotiators, but the IPCC was not able to deviate from the official line of the UNFCCC. Speaking truth to power \citep[cf.][]{Haas2004}, even revealing open secrets is not something the IPCC does.\footnote{Note the distinction between the Summary for Policy Makers and the underlying chapters. Contentious material is kept out of the summaries. The chapters contain a lot of controversial material, but are far less widely read.}

This case shows that the IPCC finds it difficult to challenge the UNFCCC.

\subsection{\textit{Ex post} policy evaluation}
Climate policy research has a long tradition of \textit{ex ante} analysis, in which a hypothetical future with a proposed policy intervention is compared to a future without. This is a key part of decision making. In the early days of climate policy research, it was the only analysis feasible: There were no actual climate policies to study.

This changed in 1990 when Poland and Finland adopted a carbon tax. Data arrive with a delay and it takes some time to write and publish a paper, but economists are creatures of habit. The first \textit{ex post} analysis of a carbon tax appeared 20 years after the first tax \citep{Lin2011}, the first analysis of a non-tax policy a year later \citep{Leahy2012}, and the first model calibration to climate policy data two years after that \citep{Tol2014En}. The IPCC can perhaps be forgiven for ignoring this nascent literature in AR5\textemdash parameters had been set before the first \textit{ex post} papers were published. In AR6, the IPCC used empirical evidence when discussing the efficacy of policy instruments and international negotiations \citep{Dubash2022IPCC, Patt2022IPCC}, but essentially ignored the econometric literature on the costs of greenhouse gas emission reduction \citep{Riahi2022IPCC}. As more and more prominent papers had been published, there is no excuse for this omission.

The IPCC's reluctance to assess the \textit{ex post} policy evaluation literature may be explained by looking at some of these papers. \citet{XING2021} find that 70\% of households who bought a subsidized electric vehicle would have bought one without the subsidy, \citet{Leahy2012} that 55\% of households did not need a subsidy to buy a biomass boiler. \citet{Fowlie2018} demonstrate that an energy efficiency programme destroyed value. \citet{Lin2011} show that a carbon tax reduced emissions in Finland, but not in Denmark, the Netherlands, Norway, and Sweden. \citet{Rafaty2020} does not find a significant effect either. However, \citet{Sen2018}, \citet{Andersson2019}, \citet{Best2020} and \citet{Metcalf2020} do find statistically significant emission reduction from carbon pricing, but did not control for carbon leakage\textemdash \citet{Aichele2015} conclude that all emission reductions under the Kyoto Protocol were offset by an increase in emissions elsewhere, \citet{Fell2018} the same for RGGI, and \citet{Koch2019} for the EU ETS\textemdash and may not have properly controlled for economic growth\textemdash \citet{Bel2015} find that lacklustre economic growth rather than climate policy accounts for the drop in carbon dioxide emissions in the EU.\footnote{None of these papers were used in the Sixth Assessment Report of Working Group III of the IPCC to validate ex-ante cost estimates \citep{IPCCWG32022}. However, Chapter 13 on policy instruments notes that ``[n]umerous assessments of specific policies, especially the EU ETS and the British Columbia carbon tax, conclude that most have reduced emissions'', claiming that the evidence is ``robust'' and the agreement ``high'' \citep{Dubash2022IPCC}.}

Some of the above studies make for uncomfortable reading for the officials and politicians who championed these policies. Academic journals only consider the quality and novelty of research. The IPCC adds another layer of review: Political approval. The IPCC does, of course, note that greenhouse gas concentrations have continued to increase but as this is a policy failure at the global scale, any individual policy maker can point the finger at her least favourite foreign counterpart. This is not the case for the policy evaluations above, all of which have a readily identifiable sponsor.

\textit{Ex post} policy evaluation highlights another problem. Although this is not necessary \citep{Barrage2019}, \textit{ex ante} studies typically consider optimal policy in a market with a single imperfection (climate change). First-best interventions are, by definition, the cheapest way to reduce emissions. \textit{Ex post} studies reveal that actual policy is a crummy third-best, and that the estimates of the costs of greenhouse gas emission reduction that can be found in IPCC reports are too optimistic \citep{Tol2014En}. Bureaucratic power is one explanation for the excessive costs \citep{Tol2020CCE}, a direct affront to the IPCC's clients.

\begin{figure}
    \centering
    \includegraphics[width=\textwidth]{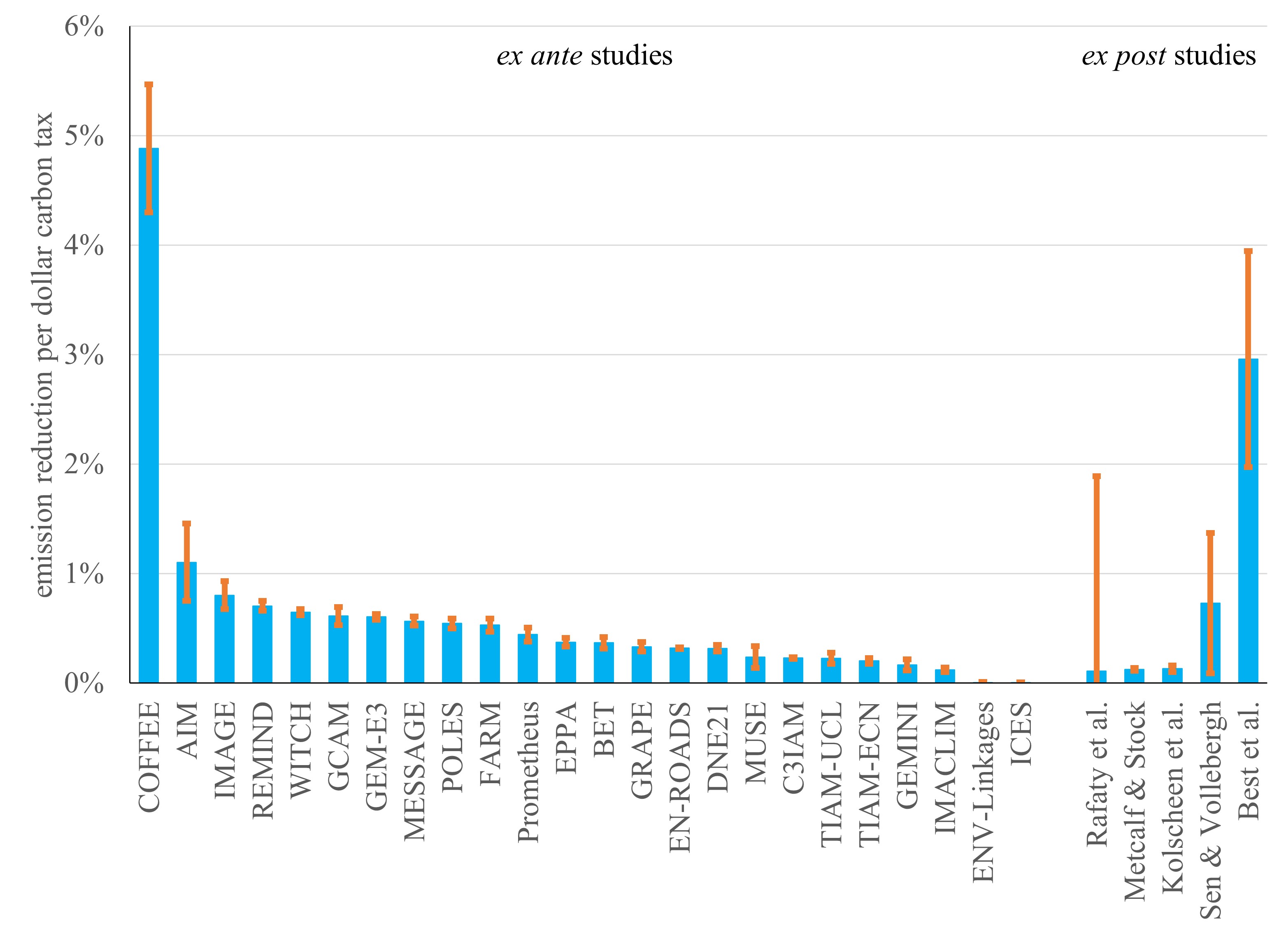}
    \caption{The efficacy of carbon pricing according to 24 \textit{ex ante} models reviewed in IPCC AR6 and 5 published \textit{ex post} studies.}
    \label{fig:expostante}
\end{figure}

\textit{Ex post} studies do not just challenge the illusions of policy makers that their interventions are effective and cheap. \text{Ex ante} modellers are challenged too. Figure \ref{fig:expostante} illustrates this. On the left, it shows results for the 24 models in the \href{https://data.ece.iiasa.ac.at/ar6/}{IPCC AR6 scenario database} that report sufficient information to compute the efficacy of a carbon tax, here defined as the percentage CO\textsuperscript{2} emission reduction (from baseline) in 2030 divided by the carbon tax (or permit price) in the 2020s. Results vary widely, from 0.0042\%/\$ for \textsc{ices} to 4.8\%/\$ for \textsc{coffee}.

On the right of Figure \ref{fig:expostante}, five \textit{ex post} estimates of the same metric are shown \citep{Rafaty2020,Kohlscheen2021,Sen2018,Metcalf2020,Best2020}. Three of these studies agree that a carbon tax of \$1/tCO\textsubscript{2} would cut emissions by some 0.12\%. The other two studies are more optimistic about the efficacy of carbon pricing. Two of the 24 \textit{ex post} models are more pessimistic, and 21 are more optimistic than these three empirical studies. Only the \textsc{imaclim} model is close to the empirical evidence.

If this analysis holds up as more econometric studies are published, then IPCC AR7 will have to confront the fact that not only policy makers were overly optimistic about climate policy, but previous IPCC reports were too.

This case reaffirms that the IPCC has difficulty admitting mistakes. It also shows that the IPCC finds it difficult to challenge the members of the Panel.

\subsection{Other issues}
The four cases above are not the only criticisms on the IPCC. Other critiques include, but are not limited to, the way uncertainties are communicated \citep{Allen2001, Patt2005, Ha-Duong2007, Risbey2007, Schenk2007, Budescu2009, Swart2009, Curry2011a, Curry2011, Mastrandrea2011, Mastrandrea2011a, Moss2011, Budescu2012, Narita2012, Sluijs2012, Adler2014, Keohane2014, Aven2015, McMahon2015, Miliauskas2016, Janzwood2020, Molina2021}, the delayed update \citep{Godal2003} and conceptualization \citep{Tol2012ERL} of global warming potentials, selection bias \citep{Tavoni2010}, inaccuracies in reporting \citep{Kintisch2010, Schiermeier2010}, conflicts of interest \citep{NewScientist2010}, and, most famous of all, the hockey stick \citep{McIntyre2003, McIntyre2005, Brumfiel2006, Holland2007, Frank2010, Wegman2010, Mann2012, Montford2012, Montford2015, Zorita2019, Mann2021}. In many of these episodes, the IPCC comes across as a bureaucracy defending its reputation. In academia, of course, a reputation is best maintained by quickly acknowledging and correcting mistakes.

\section{Remedies}
\label{sc:remedies}
A common thread in the last three case studies above is that the IPCC finds it difficult to tell policy makers that they were wrong. The first and fourth case study reveals that the IPCC also finds it difficult to admit that it was wrong itself. For the IPCC to change and start speaking truth to power, it first needs to admit that it has found it difficult in the past to do so. Reforming the IPCC will therefore not be easy.

Abolishing the IPCC is not an option. The international negotiations on climate policy would be more fraught and make less progress without a common knowledge base. Negotiations would quickly move from disagreements about what to do to disagreements about what is what. Replacing the IPCC is not an option either. It is a natural monopoly that took years to build and that has successfully erected barriers to entry \citep{Peiser2007, Tol2011CC}. IPCC reform is the only feasible way forward.

Reforming the IPCC is necessary but not easy. It is not even clear that the IPCC realizes its need to reform. Climate policy has a long traditions of good intentions and poor delivery. Many targets have been announced and missed, but policy failure has typically been obscured by the announcement of more ambitious targets for a later date. The first stock-take under the Paris Agreement can, and almost certainly will be fudged\textemdash emissions are not falling as fast as planned, but we can make up for the shortfall later\textemdash but the 1.5\celsius{} target will be breached sooner rather than later.

Climate policy has been effective. Renewable energy is cheaper than most expected two decades ago, electrification of transport and heating is easier than was feared, and novel protein foods make unexpectedly fast inroads. In time, greenhouse gas emissions will fall. But politicians promised much more than that, and those promises will come to haunt them before AR7 is published.

The first global stock-take presents an opportunity to test a new mode of operation for the IPCC. Updates of data and projections of greenhouse gas emissions are now routinely accompanied by assessments of their implications using simple climate models. Such activities could be brought under the IPCC umbrella. Furthermore, new observations on emissions also tell us something about the relative success or failure of past climate policies. The IPCC could assess this as replication-with-an-extended-dataset is a routine activity, rather than new research.

But the IPCC should go further. A relatively simple change is to the structure of the IPCC assessment reports. There are now four: Science, Impacts, Mitigation, and Synthesis. The Synthesis Report does not add anything new, instead syntheses the first three reports. The Synthesis Report could be extended with a few chapters that assess the academic literature that synthesize climate research. Particularly, papers that compare the costs of greenhouse gas emission reduction to the avoided impacts of climate change fit neither in Working Group II nor in Working Group III, but should inform the Synthesis Report.

The key problem with the IPCC is that the panel consists of government representatives. Depending on the country, lead representatives are either from the state department or from the environment agency. Their interest is political, not academic. This was probably a good design in the early IPCC years. Political buy-in was critical then. It has been won. The IPCC will now need to show a more independent streak, of course without losing government support.

Many governments have parts that operate somewhat outside political control. Central banks are one example, budget watchdogs another, as are inspectorates and ombudspeople. Control of the IPCC could be transferred to an organization like that. However, it may be more appropriate to put either the departments of higher education, research funding agencies, or even the national academies of science in charge. This would have three advantages. First, it would make the IPCC less ``closed-off'' \citep{Hermansen2021}. Second, it would put academic rigour before political expedience. Third, researchers would be nominated to the IPCC on the basis of their academic credentials. More independent-minded authors and a less politicized panel would allow the IPCC to be more critical of its past and more frank about policy. Fourth, changes at the Panel would also lead to changes at the IPCC Bureau, which is currently dominated by apparatchiks. The IPCC should instead be lead by more visionary people, as it indeed was in the beginning.

A more academic IPCC would allow it to discuss controversial topics like monetary valuation, cost-benefit analysis and game theory. These topics should be discussed as they are, that is, tools for analysis with advantages and disadvantages. Its conclusions of which should not be taken as gospel but instead be used to inform and enlighten. Its recommendations should be not be seen as prescriptions but as suggestions for improvement.

A more academic IPCC would also attract more top scholars. Top scholars in the climate sciences are in WG1 of the IPCC, but WG2 and WG3 have been less successful in this regard. Top people would be not as beholden to the political forces in the IPCC.

\section{Discussion and conclusion}
\label{sc:conclude}
The IPCC has played a key role in national and international climate policy, convincing people that the arguments for greenhouse gas emission reduction are sound and providing a common knowledge base that allows policy discussions to focus on solutions. But the policy landscape is changing. The key question is no longer ``should we do something?''. The main issue is now ``are we doing the right thing?'' If the IPCC wants to live up to its mantra ``always policy relevant, never policy prescriptive'', it should change too.

The case studies presented in this paper suggest that the IPCC is an organization that finds it difficult to admit problems and the need to change; one of the IPCC's key problems is its unwillingness or inability to present politically inexpedient results. Academic freedom was instituted for exactly this reason. Professors should present their results regardless of how the chips may fall. The IPCC is not an academic organization, however. Making it one would reduce the political buy-in that is crucial to its functioning. I therefore propose to shift control over the IPCC to those parts of government that are closest to academia.

The IPCC has a long history of ignoring what I have to say\textemdash some people in the organization may even take pride in that. But the truth of the matter is that the IPCC has not changed that much in its 35-year existence whereas climate policy has changed beyond recognition. The IPCC will have to change too, lest it sails into oblivion.

\bibliography{master}

\begin{thebibliography}{150}
\providecommand{\natexlab}[1]{#1}
\providecommand{\url}[1]{\texttt{#1}}
\expandafter\ifx\csname urlstyle\endcsname\relax
  \providecommand{\doi}[1]{doi: #1}\else
  \providecommand{\doi}{doi: \begingroup \urlstyle{rm}\Url}\fi

\bibitem[Adler and Hirsch~Hadorn(2014)]{Adler2014}
C.~Adler and G.~Hirsch~Hadorn.
\newblock The {IPCC} and treatment of uncertainties: Topics and sources of
  dissensus.
\newblock \emph{Wiley Interdisciplinary Reviews: Climate Change}, 5\penalty0
  (5):\penalty0 663--676, 2014.
\newblock \doi{10.1002/wcc.297}.

\bibitem[Agrawala(1998{\natexlab{a}})]{Agrawala1998a}
S.~Agrawala.
\newblock Context and early origins of the {Intergovernmental Panel on Climate
  Change}.
\newblock \emph{Climatic Change}, 39\penalty0 (4):\penalty0 605--620,
  1998{\natexlab{a}}.
\newblock \doi{10.1023/A:1005315532386}.

\bibitem[Agrawala(1998{\natexlab{b}})]{Agrawala1998b}
S.~Agrawala.
\newblock Structural and process history of the {Intergovernmental Panel on
  Climate Change}.
\newblock \emph{Climatic Change}, 39\penalty0 (4):\penalty0 621--642,
  1998{\natexlab{b}}.
\newblock \doi{10.1023/A:1005312331477}.

\bibitem[Aichele and Felbermayr(2015)]{Aichele2015}
R.~Aichele and G.~Felbermayr.
\newblock {Kyoto and Carbon Leakage: An Empirical Analysis of the Carbon
  Content of Bilateral Trade}.
\newblock \emph{The Review of Economics and Statistics}, 97\penalty0
  (1):\penalty0 104--115, 2015.

\bibitem[Alexander(2007)]{Alexander2007}
W.~J.~R. Alexander.
\newblock The {IPCC}: Structure, processes and politics climate
  change\textemdash the failure of science.
\newblock \emph{Energy and Environment}, 18\penalty0 (7-8):\penalty0
  1073--1077, 2007.
\newblock \doi{10.1260/095830507782616805}.

\bibitem[Allen et~al.(2001)Allen, Raper, and Mitchell]{Allen2001}
M.~Allen, S.~Raper, and J.~Mitchell.
\newblock Uncertainty in the {IPCC}'s {Third Assessment Report}.
\newblock \emph{Science}, 293\penalty0 (5529):\penalty0 430+431+433, 2001.

\bibitem[Andersson(2019)]{Andersson2019}
J.~J. Andersson.
\newblock {Carbon Taxes and CO\textsubscript{2} Emissions: Sweden as a Case
  Study}.
\newblock \emph{American Economic Journal: Economic Policy}, 11\penalty0
  (4):\penalty0 1--30, 2019.

\bibitem[Arent et~al.(2014)Arent, Tol, Faust, Hella, Kumar, Strzepek, Toth, and
  Yan]{Arent2014}
D.~Arent, R.~S.~J. Tol, E.~Faust, J.~P. Hella, S.~Kumar, K.~M. Strzepek, F.~L.
  Toth, and D.~Yan.
\newblock Key economic sectors and services.
\newblock In C.~Field, V.~Barros, D.~Dokken, K.~Mach, M.~Mastrandrea, T.~Bilir,
  M.~Chatterjee, K.~Ebi, Y.~Estrada, R.~Genova, B.~Girma, E.~Kissel, A.~Levy,
  S.~MacCracken, P.~Mastrandrea, and L.~White, editors, \emph{Climate Change
  2014: Impacts, Adaptation, and Vulnerability. Part A: Global and Sectoral
  Aspects. Contribution of Working Group II to the Fifth Assessment Report of
  the Intergovernmental Panel on Climate Change}, chapter~10, pages 659--708.
  Cambridge University Press, Cambridge, 2014.

\bibitem[Arrow et~al.(1996)Arrow, Cline, Maler, Munasinghe, SQUITIERI, and
  Stiglitz]{Arrow1996IPCC}
K.~J. Arrow, W.~R. Cline, K.-G. Maler, M.~Munasinghe, R.~SQUITIERI, and J.~E.
  Stiglitz.
\newblock Intertemporal equity, discounting, and economic efficiency.
\newblock In J.~P. Bruce, H.~Lee, and E.~F. Haites, editors, \emph{Climate
  Change 1995: Economic and Social Dimensions of Climate Change\textemdash
  Contribution of Working Group III to the Second Assessment Report of the
  Intergovernmental Panel on Climate Change}, chapter~4. Cambridge University
  Press, Cambridge, 1996.

\bibitem[Aven and Renn(2015)]{Aven2015}
T.~Aven and O.~Renn.
\newblock An evaluation of the treatment of risk and uncertainties in the
  {IPCC} reports on climate change.
\newblock \emph{Risk Analysis}, 35\penalty0 (4):\penalty0 701--712, 2015.
\newblock \doi{10.1111/risa.12298}.

\bibitem[Barrage(2019)]{Barrage2019}
L.~Barrage.
\newblock {Optimal Dynamic Carbon Taxes in a Climate–Economy Model with
  Distortionary Fiscal Policy}.
\newblock \emph{The Review of Economic Studies}, 87\penalty0 (1):\penalty0
  1--39, 2019.
\newblock URL \url{https://doi.org/10.1093/restud/rdz055}.

\bibitem[Barrett(1994)]{Barrett1994}
S.~Barrett.
\newblock Self-enforcing international environmental agreements.
\newblock \emph{Oxford Economic Papers}, 46:\penalty0 878--894, 1994.
\newblock \doi{10.1093/oep/46.Supplement_1.878}.

\bibitem[Barro and Sala-I-Martin(1992)]{Barro1992}
R.~J. Barro and X.~Sala-I-Martin.
\newblock Convergence.
\newblock \emph{Journal of Political Economy}, pages 223--251, 1992.
\newblock \doi{10.1086/261816}.

\bibitem[Battaglini and Harstad(2016)]{Battaglini2016}
M.~Battaglini and B.~Harstad.
\newblock Participation and duration of environmental agreements.
\newblock \emph{Journal of Political Economy}, 124\penalty0 (1):\penalty0
  160--204, 2016.
\newblock \doi{10.1086/684478}.

\bibitem[Bel and Joseph(2015)]{Bel2015}
G.~Bel and S.~Joseph.
\newblock Emission abatement: Untangling the impacts of the eu ets and the
  economic crisis.
\newblock \emph{Energy Economics}, 49:\penalty0 531--539, 2015.
\newblock ISSN 0140-9883.
\newblock \doi{http://dx.doi.org/10.1016/j.eneco.2015.03.014}.

\bibitem[Best et~al.(2020)Best, Burke, and Jotzo]{Best2020}
R.~Best, P.~J. Burke, and F.~Jotzo.
\newblock {Carbon Pricing Efficacy: Cross-Country Evidence}.
\newblock \emph{Environmental \& Resource Economics}, 77\penalty0 (1):\penalty0
  69--94, 2020.
\newblock \doi{10.1007/s10640-020-00436-}.

\bibitem[Bradford(2008)]{Bradford2008}
D.~F. Bradford.
\newblock Improving on kyoto: Greenhouse gas control as the purchase of a
  global public good.
\newblock In R.~Guesnerie and H.~Tulkens, editors, \emph{The Design of Climate
  Policy}, chapter~2, pages 12--36. MIT Press, Cambridge, 2008.
\newblock \doi{https://doi.org/10.7551/mitpress/9780262073028.003.0002}.

\bibitem[Bruce(1995)]{Bruce1995}
J.~P. Bruce.
\newblock Impact of climate change.
\newblock \emph{Nature}, 377\penalty0 (6549):\penalty0 472, 1995.
\newblock \doi{10.1038/377472a0}.

\bibitem[Brumfiel(2006)]{Brumfiel2006}
G.~Brumfiel.
\newblock Academy affirms hockey-stick graph.
\newblock \emph{Nature}, 441\penalty0 (7097):\penalty0 1032--1033, 2006.
\newblock \doi{10.1038/4411032a}.

\bibitem[Budescu et~al.(2009)Budescu, Broomell, and Por]{Budescu2009}
D.~Budescu, S.~Broomell, and H.-H. Por.
\newblock Improving communication of uncertainty in the reports of the
  {Intergovernmental Panel on Climate Change}.
\newblock \emph{Psychological Science}, 20\penalty0 (3):\penalty0 299--308,
  2009.
\newblock \doi{10.1111/j.1467-9280.2009.02284.x}.

\bibitem[Budescu et~al.(2012)Budescu, Por, and Broomell]{Budescu2012}
D.~V. Budescu, H.-H. Por, and S.~B. Broomell.
\newblock Effective communication of uncertainty in the {IPCC} reports.
\newblock \emph{Climatic Change}, 113\penalty0 (2):\penalty0 181--200, 2012.
\newblock \doi{10.1007/s10584-011-0330-3}.

\bibitem[Burgess et~al.(2020)Burgess, Ritchie, Shapland, and {Pielke
  Jr}]{Burgess2020}
M.~G. Burgess, J.~Ritchie, J.~Shapland, and R.~A. {Pielke Jr}.
\newblock {IPCC} baseline scenarios have over-projected co\textsubscript{2}
  emissions and economic growth.
\newblock \emph{Environmental Research Letters}, 16\penalty0 (1):\penalty0
  024027, 2020.
\newblock \doi{10.1088/1748-9326/abcdd2}.

\bibitem[Carraro and Siniscalco(1992)]{Carraro1992}
C.~Carraro and D.~Siniscalco.
\newblock The international dimension of environmental policy.
\newblock \emph{European Economic Review}, 36\penalty0 (2-3):\penalty0
  379--387, 1992.
\newblock \doi{10.1016/0014-2921(92)90094-D}.

\bibitem[Carraro and Siniscalco(1993)]{Carraro1993}
C.~Carraro and D.~Siniscalco.
\newblock Strategies for the international protection of the environment.
\newblock \emph{Journal of Public Economics}, 52\penalty0 (3):\penalty0
  309--328, 1993.
\newblock \doi{10.1016/0047-2727(93)90037-T}.

\bibitem[Carraro et~al.(2015)Carraro, Edenhofer, Flachsland, Kolstad, Stavins,
  and Stowe]{Carraro2015}
C.~Carraro, O.~Edenhofer, C.~Flachsland, C.~Kolstad, R.~Stavins, and R.~Stowe.
\newblock The {IPCC} at a crossroads: Opportunities for reform.
\newblock \emph{Science}, 350\penalty0 (6256):\penalty0 34--35, 2015.
\newblock \doi{10.1126/science.aac4419}.

\bibitem[Castles and Henderson(2003{\natexlab{a}})]{Castles2003a}
I.~Castles and D.~Henderson.
\newblock The {IPCC} emission scenarios: An economic-statistical critique.
\newblock \emph{Energy and Environment}, 14\penalty0 (2-3):\penalty0 159--186,
  2003{\natexlab{a}}.
\newblock \doi{10.1260/095830503765184583}.

\bibitem[Castles and Henderson(2003{\natexlab{b}})]{Castles2003b}
I.~Castles and D.~Henderson.
\newblock Economics, emissions scenarios and the work of the {IPCC}.
\newblock \emph{Energy and Environment}, 14\penalty0 (4):\penalty0 415--436,
  2003{\natexlab{b}}.
\newblock \doi{10.1260/095830503322364430}.

\bibitem[Chan et~al.(2016)Chan, Carraro, Edenhofer, Kolstad, and
  Stavins]{Chan2016}
G.~Chan, C.~Carraro, O.~Edenhofer, C.~Kolstad, and R.~Stavins.
\newblock Reforming the {IPCC}'s assessment of climate change economics.
\newblock \emph{Climate Change Economics}, 7\penalty0 (1), 2016.
\newblock \doi{10.1142/S2010007816400017}.

\bibitem[Cointe et~al.(2019)Cointe, Cassen, and Nada\"{i}]{Cointe2019}
B.~Cointe, C.~Cassen, and A.~Nada\"{i}.
\newblock Organising policy-relevant knowledge for climate action: Integrated
  assessment modelling, the {IPCC}, and the emergence of a collective expertise
  on socioeconomic emission scenarios.
\newblock \emph{Science \& Technology Studies}, 32\penalty0 (4):\penalty0
  36–57, 2019.
\newblock \doi{10.23987/sts.65031}.
\newblock URL
  \url{https://sciencetechnologystudies.journal.fi/article/view/65031}.

\bibitem[Courtney(1996)]{Courtney1996}
R.~S. Courtney.
\newblock Purpose and function of {IPCC}.
\newblock \emph{Nature}, 379\penalty0 (6561):\penalty0 109, 1996.
\newblock \doi{10.1038/379109b0}.

\bibitem[Curry(2011)]{Curry2011a}
J.~Curry.
\newblock Reasoning about climate uncertainty.
\newblock \emph{Climatic Change}, 108\penalty0 (4):\penalty0 723--732, 2011.
\newblock \doi{10.1007/s10584-011-0180-z}.

\bibitem[Curry and Webster(2011)]{Curry2011}
J.~Curry and P.~Webster.
\newblock Climate science and the uncertainty monster.
\newblock \emph{Bulletin of the American Meteorological Society}, 92\penalty0
  (12):\penalty0 1667--1682, 2011.
\newblock \doi{10.1175/2011BAMS3139.1}.

\bibitem[Dawson(1999)]{Dawson1999}
L.~L. Dawson.
\newblock When prophecy fails and faith persists: A theoretical overview.
\newblock \emph{Nova Religio: The Journal of Alternative and Emergent
  Religions}, 3\penalty0 (1):\penalty0 60--82, 1999.
\newblock ISSN 10926690, 15418480.
\newblock URL \url{http://www.jstor.org/stable/10.1525/nr.1999.3.1.60}.

\bibitem[Dein(2001)]{Dein2001}
S.~Dein.
\newblock What really happens when prophecy fails: The case of {L}ubavitch.
\newblock \emph{Sociology of Religion}, 62\penalty0 (3):\penalty0 383--401,
  2001.
\newblock ISSN 10694404, 17598818.
\newblock URL \url{http://www.jstor.org/stable/3712356}.

\bibitem[Duflo(2017)]{Duflo2017}
E.~Duflo.
\newblock The economist as plumber.
\newblock \emph{American Economic Review}, 107\penalty0 (5):\penalty0 1--26,
  May 2017.
\newblock URL \url{https://www.aeaweb.org/articles?id=10.1257/aer.p20171153}.

\bibitem[Edwards and Schneider(2001)]{Edwards2001}
P.~N. Edwards and S.~H. Schneider.
\newblock Self-governance and peer review in science-for-policy: The case of
  the {{IPCC} Second Assessment Report}.
\newblock In C.~Miller and P.~N. Edwards, editors, \emph{Changing the
  Atmosphere: Expert Knowledge and Environmental Governance}, chapter~7, pages
  219--246. MIT Press, Cambridge, 2001.

\bibitem[Fankhauser(1997)]{Fankhauser1997b}
S.~Fankhauser.
\newblock The social costs of climate change: The {IPCC Second Assessment
  Report} and beyond.
\newblock \emph{Mitigation and Adaptation Strategies for Global Change},
  1\penalty0 (4):\penalty0 385--403, 1997.
\newblock \doi{10.1007/bf00464889}.

\bibitem[Fankhauser et~al.(1998)Fankhauser, Tol, and Pearce]{Fankhauser1998}
S.~Fankhauser, R.~S.~J. Tol, and D.~W. Pearce.
\newblock Extensions and alternatives to climate change impact valuation: On
  the critique of {IPCC Working Group III}'s impact estimates.
\newblock \emph{Environment and Development Economics}, 3\penalty0
  (1):\penalty0 59--81, 1998.
\newblock \doi{10.1017/S1355770X98000047}.

\bibitem[Fell and Maniloff(2018)]{Fell2018}
H.~Fell and P.~Maniloff.
\newblock {Leakage in regional environmental policy: The case of the regional
  greenhouse gas initiative}.
\newblock \emph{Journal of Environmental Economics and Management}, 87\penalty0
  (C):\penalty0 1--23, 2018.
\newblock \doi{10.1016/j.jeem.2017.10.00}.

\bibitem[Festinger et~al.(1956)Festinger, Riecken, and
  Schachter]{Festinger1956}
L.~Festinger, H.~W. Riecken, and S.~Schachter.
\newblock \emph{When prophecy fails\textemdash A social and psychological study
  of a modern group that predicted the destruction of the world}.
\newblock University of Minnesota Press, 1956.

\bibitem[Fowlie et~al.(2018)Fowlie, Greenstone, and Wolfram]{Fowlie2018}
M.~Fowlie, M.~Greenstone, and C.~Wolfram.
\newblock Do energy efficiency investments deliver? evidence from the
  weatherization assistance program*.
\newblock \emph{The Quarterly Journal of Economics}, 133\penalty0 (3):\penalty0
  1597--1644, 2018.
\newblock \doi{10.1093/qje/qjy005}.

\bibitem[Frank et~al.(2010)Frank, Esper, Zorita, and Wilson]{Frank2010}
D.~Frank, J.~Esper, E.~Zorita, and R.~Wilson.
\newblock A noodle, hockey stick, and spaghetti plate: A perspective on
  high-resolution paleoclimatology.
\newblock \emph{Wiley Interdisciplinary Reviews: Climate Change}, 1\penalty0
  (4):\penalty0 507--516, 2010.
\newblock \doi{10.1002/wcc.53}.

\bibitem[Girod et~al.(2009)Girod, Wiek, Mieg, and Hulme]{Girod2009}
B.~Girod, A.~Wiek, H.~Mieg, and M.~Hulme.
\newblock The evolution of the {IPCC}'s emissions scenarios.
\newblock \emph{Environmental Science and Policy}, 12\penalty0 (2):\penalty0
  103--118, 2009.
\newblock \doi{10.1016/j.envsci.2008.12.006}.

\bibitem[Godal(2003)]{Godal2003}
O.~Godal.
\newblock The {IPCC}'s assessment of multidisciplinary issues: The case of
  greenhouse gas indices. an editorial essay.
\newblock \emph{Climatic Change}, 58\penalty0 (3):\penalty0 24--249, 2003.
\newblock \doi{10.1023/A:1023935918891}.

\bibitem[Grubb(1996)]{Grubb1996}
M.~Grubb.
\newblock Purpose and function of {IPCC}.
\newblock \emph{Nature}, 379\penalty0 (6561):\penalty0 108, 1996.
\newblock \doi{10.1038/379108a0}.

\bibitem[Gr\"{u}bler et~al.(2004)Gr\"{u}bler, Nakicenovic, Alcamo, Davis,
  Fenhann, Hare, Mori, Pepper, Pitcher, Riahi, Rogner, La~Rovere, Sankovski,
  Schlesinger, Shukla, Swart, Victor, and Jung]{Grubler2004}
A.~Gr\"{u}bler, N.~Nakicenovic, J.~Alcamo, G.~Davis, J.~Fenhann, B.~Hare,
  S.~Mori, B.~Pepper, H.~Pitcher, K.~Riahi, H.-H. Rogner, E.~La~Rovere,
  A.~Sankovski, M.~Schlesinger, R.~Shukla, R.~Swart, N.~Victor, and T.~Jung.
\newblock Emissions scenarios: A final response.
\newblock \emph{Energy and Environment}, 15\penalty0 (1):\penalty0 11--24,
  2004.
\newblock \doi{10.1260/095830504322986466}.

\bibitem[Ha-Duong et~al.(2007)Ha-Duong, Swart, Bernstein, and
  Petersen]{Ha-Duong2007}
M.~Ha-Duong, R.~Swart, L.~Bernstein, and A.~Petersen.
\newblock Uncertainty management in the {IPCC}: Agreeing to disagree.
\newblock \emph{Global Environmental Change}, 17\penalty0 (1):\penalty0 8--11,
  2007.
\newblock \doi{10.1016/j.gloenvcha.2006.12.003}.

\bibitem[Haas(2004)]{Haas2004}
P.~Haas.
\newblock When does power listen to truth? a constructivist approach to the
  policy process.
\newblock \emph{Journal of European Public Policy}, 11\penalty0 (4):\penalty0
  569--592, 2004.
\newblock URL \url{https://doi.org/10.1080/1350176042000248034}.

\bibitem[H\"{a}nsel et~al.(2020)H\"{a}nsel, Drupp, Johansson, Nesje, Azar,
  Freeman, Groom, and Sterner]{Hansel2020}
M.~C. H\"{a}nsel, M.~A. Drupp, D.~J.~A. Johansson, F.~Nesje, C.~Azar, M.~C.
  Freeman, B.~Groom, and T.~Sterner.
\newblock Climate economics support for the {UN} climate targets.
\newblock \emph{Nature Climate Change}, 10\penalty0 (8):\penalty0 781--789,
  2020.
\newblock \doi{10.1038/s41558-020-0833-x}.

\bibitem[Hermansen et~al.(2021)Hermansen, Lahn, Sundqvist, and
  \O~ye]{Hermansen2021}
E.~A.~T. Hermansen, B.~Lahn, G.~Sundqvist, and E.~\O~ye.
\newblock Post-{P}aris policy relevance: lessons from the {IPCC SR15} process.
\newblock \emph{Climatic Change}, 169\penalty0 (1-2), 2021.
\newblock \doi{10.1007/s10584-021-03210-0}.

\bibitem[Holland(2007)]{Holland2007}
D.~Holland.
\newblock Bias and concealment in the {IPCC} process: The "hockey-stick" affair
  and its implications.
\newblock \emph{Energy and Environment}, 18\penalty0 (7-8):\penalty0 951--983,
  2007.
\newblock \doi{10.1260/095830507782616788}.

\bibitem[Holtsmark and Alfsen(2004)]{Holtsmark2004}
B.~J. Holtsmark and K.~H. Alfsen.
\newblock The use of {PPP} or {MER} in the construction of emission scenarios
  is more than a question of ‘metrics’.
\newblock \emph{Climate Policy}, 4\penalty0 (2):\penalty0 205--216, 2004.
\newblock \doi{10.1080/14693062.2004.9685521}.

\bibitem[Holtsmark and Alfsen(2005)]{Holtsmark2005}
B.~J. Holtsmark and K.~H. Alfsen.
\newblock {PPP} correction of the {IPCC} emission scenarios - does it matter?
\newblock \emph{Climatic Change}, 68\penalty0 (1-2):\penalty0 11--19, 2005.
\newblock \doi{10.1007/s10584-005-1310-2}.

\bibitem[Hulme and Mahony(2010)]{Hulme2010}
M.~Hulme and M.~Mahony.
\newblock Climate change: What do we know about the {IPCC}?
\newblock \emph{Progress in Physical Geography}, 34\penalty0 (5):\penalty0
  705--718, 2010.
\newblock \doi{10.1177/0309133310373719}.

\bibitem[Janzwood(2020)]{Janzwood2020}
S.~Janzwood.
\newblock Confident, likely, or both? the implementation of the uncertainty
  language framework in {IPCC} special reports.
\newblock \emph{Climatic Change}, 162\penalty0 (3):\penalty0 1655--1675, 2020.
\newblock \doi{10.1007/s10584-020-02746-x}.

\bibitem[Keohane et~al.(2014)Keohane, Lane, and Oppenheimer]{Keohane2014}
R.~O. Keohane, M.~Lane, and M.~Oppenheimer.
\newblock The ethics of scientific communication under uncertainty.
\newblock \emph{Politics, Philosophy and Economics}, 13\penalty0 (4):\penalty0
  343--368, 2014.
\newblock \doi{10.1177/1470594X14538570}.

\bibitem[Kintisch(2010)]{Kintisch2010}
E.~Kintisch.
\newblock {IPCC} seeks 'broader community engagement' to correct errors.
\newblock \emph{Science}, 327\penalty0 (5967):\penalty0 768--769, 2010.
\newblock \doi{10.1126/science.327.5967.768-b}.

\bibitem[Koch and Basse~Mama(2019)]{Koch2019}
N.~Koch and H.~Basse~Mama.
\newblock {Does the {EU} Emissions Trading System induce investment leakage?
  Evidence from {G}erman multinational firms}.
\newblock \emph{Energy Economics}, 81\penalty0 (C):\penalty0 479--492, 2019.
\newblock \doi{10.1016/j.eneco.2019.04.0}.

\bibitem[Kohlscheen et~al.(2021)Kohlscheen, Moessner, and
  Tak\'{a}ts]{Kohlscheen2021}
E.~Kohlscheen, R.~Moessner, and E.~Tak\'{a}ts.
\newblock Effects of carbon pricing and other climate policies on
  {CO}\textsubscript{2} emissions.
\newblock Working Paper 9347, {CESifo}, 2021.
\newblock URL
  \url{https://www.cesifo.org/en/publikationen/2021/working-paper/effects-carbon-pricing-and-other-climate-policies-co2-emissions}.

\bibitem[K. Dubash et~al.(2022)K. Dubash, Mitchell, Boasson,
  J. Borbor-Córdova, Fifita, Erik Haites, Jaccard, Jotzo, Naidoo,
  Romero-Lankao, Shen, Shlapak, and Wu]{Dubash2022IPCC}
N.~K. Dubash, C.~Mitchell, E.~L. Boasson, M.~J. Borbor-Córdova, S.~Fifita,
  Erik Haites, M.~Jaccard, F.~Jotzo, S.~Naidoo, P.~Romero-Lankao, W.~Shen,
  M.~Shlapak, and L.~Wu.
\newblock National and sub-national policies and institutions.
\newblock In P.~R. Shukla, J.~Skea, R.~Slade, A.~A. Khourdajie, R.~van Diemen,
  D.~McCollum, M.~Pathak, S.~Some, P.~Vyas, R.~Fradera, M.~Belkacemi,
  A.~Hasija, G.~Lisboa, S.~Luz, and J.~Malley, editors, \emph{Climate Change
  2022: Mitigation of Climate Change\textemdash Contribution of Working Group
  III to the Sixth Assessment Report of the Intergovernmental Panel on Climate
  Change}. Cambridge University Press, Cambridge, 2022.

\bibitem[Leahy and Tol(2012)]{Leahy2012}
E.~Leahy and R.~S.~J. Tol.
\newblock {Greener homes: an ex-post estimate of the cost of carbon dioxide
  emission reduction using administrative micro-data from the Republic of
  Ireland}.
\newblock \emph{Environmental Economics and Policy Studies}, 14\penalty0
  (3):\penalty0 219--239, 2012.
\newblock \doi{10.1007/s10018-012-0034-6}.

\bibitem[Leggett et~al.(1992)Leggett, Pepper, and Swart]{Leggett1992}
J.~Leggett, W.~Pepper, and R.~Swart.
\newblock Emissions scenarios for the {IPCC}: An update.
\newblock In J.~T. Houghton, B.~A. Callander, and S.~K. Varney, editors,
  \emph{Climate Change 1992: The Supplementary Report to the {IPCC} Scientific
  Assessment}, chapter~A3, pages 69--95. Cambridge University Press, Cambridge,
  1992.

\bibitem[Lin and Li(2011)]{Lin2011}
B.~Lin and X.~Li.
\newblock {The effect of carbon tax on per capita CO\textsubscript{2}
  emissions}.
\newblock \emph{Energy Policy}, 39\penalty0 (9):\penalty0 5137--5146, 2011.

\bibitem[Mann(2021)]{Mann2021}
M.~Mann.
\newblock Beyond the hockey stick: Climate lessons from the common era.
\newblock \emph{Proceedings of the National Academy of Sciences of the United
  States of America}, 118\penalty0 (39), 2021.
\newblock \doi{10.1073/PNAS.2112797118}.

\bibitem[Mann(2012)]{Mann2012}
M.~E. Mann.
\newblock \emph{The Hockey Stick and the Climate Wars: Dispatches from the
  Front Lines}.
\newblock Columbia University Press, New York, 2012.

\bibitem[Manne et~al.(2005)Manne, Richels, and Edmonds]{Manne2005}
A.~S. Manne, R.~G. Richels, and J.~A. Edmonds.
\newblock Market exchange rates or purchasing power parity: Does the choice
  make a difference to the climate debate?
\newblock \emph{Climatic Change}, 71\penalty0 (1-2):\penalty0 1--8, 2005.
\newblock \doi{10.1007/s10584-005-0470-4}.

\bibitem[Masood(1995)]{Masood1995a}
E.~Masood.
\newblock Developing countries dispute use of figures on climate change
  impacts.
\newblock \emph{Nature}, 376\penalty0 (6539):\penalty0 374, 1995.
\newblock \doi{10.1038/376374b0}.

\bibitem[Masood and Ochert(1995)]{Masood1995b}
E.~Masood and A.~Ochert.
\newblock Un climate change report turns up the heat.
\newblock \emph{Nature}, 378\penalty0 (6553):\penalty0 119, 1995.
\newblock \doi{10.1038/378119a0}.

\bibitem[Mastrandrea and Mach(2011)]{Mastrandrea2011a}
M.~D. Mastrandrea and K.~J. Mach.
\newblock Treatment of uncertainties in {IPCC} assessment reports: Past
  approaches and considerations for the {Fifth Assessment Report}.
\newblock \emph{Climatic Change}, 108\penalty0 (4):\penalty0 659--673, 2011.
\newblock \doi{10.1007/s10584-011-0177-7}.

\bibitem[Mastrandrea et~al.(2011)Mastrandrea, Mach, Plattner, Edenhofer,
  Stocker, Field, Ebi, and Matschoss]{Mastrandrea2011}
M.~D. Mastrandrea, K.~J. Mach, G.-K. Plattner, O.~Edenhofer, T.~F. Stocker,
  C.~B. Field, K.~L. Ebi, and P.~R. Matschoss.
\newblock The {IPCC} {AR5} guidance note on consistent treatment of
  uncertainties: A common approach across the working groups.
\newblock \emph{Climatic Change}, 108\penalty0 (4):\penalty0 675--691, 2011.
\newblock \doi{10.1007/s10584-011-0178-6}.

\bibitem[McIntyre and McKitrick(2003)]{McIntyre2003}
S.~McIntyre and R.~McKitrick.
\newblock Corrections to the {M}ann et al. (1998) proxy data base and northern
  hemispheric average temperature series.
\newblock \emph{Energy and Environment}, 14\penalty0 (6):\penalty0 751--772,
  2003.
\newblock \doi{10.1260/095830503322793632}.

\bibitem[McIntyre and McKitrick(2005)]{McIntyre2005}
S.~McIntyre and R.~McKitrick.
\newblock Hockey sticks, principal components, and spurious significance.
\newblock \emph{Geophysical Research Letters}, 32\penalty0 (3):\penalty0 1--5,
  2005.
\newblock \doi{10.1029/2004GL021750}.

\bibitem[McMahon et~al.(2015)McMahon, Stauffacher, and Knutti]{McMahon2015}
R.~McMahon, M.~Stauffacher, and R.~Knutti.
\newblock The unseen uncertainties in climate change: reviewing comprehension
  of an {IPCC} scenario graph.
\newblock \emph{Climatic Change}, 133\penalty0 (2):\penalty0 141--154, 2015.
\newblock \doi{10.1007/s10584-015-1473-4}.

\bibitem[Melton(1985)]{Melton1985}
J.~G. Melton.
\newblock spiritualization and reaffirmation: what really happens when prophecy
  fails.
\newblock \emph{American Studies}, 26\penalty0 (2):\penalty0 17--29, 1985.
\newblock ISSN 00263079, 21536856.
\newblock URL \url{http://www.jstor.org/stable/40641958}.

\bibitem[Metcalf and Stock(2020)]{Metcalf2020}
G.~E. Metcalf and J.~H. Stock.
\newblock The macroeconomic impact of {E}urope’s carbon taxes.
\newblock Working Paper 27488, National Bureau of Economic Research, 2020.

\bibitem[Meyer(1995{\natexlab{a}})]{Meyer1995a}
A.~Meyer.
\newblock Costing calamity.
\newblock \emph{New Scientist}, page~64, 1995{\natexlab{a}}.

\bibitem[Meyer(1995{\natexlab{b}})]{Meyer1995b}
A.~Meyer.
\newblock Economics of climate change.
\newblock \emph{Nature}, 378\penalty0 (6556):\penalty0 433, 1995{\natexlab{b}}.
\newblock \doi{10.1038/378433a0}.

\bibitem[Miliauskas and Anderson(2016)]{Miliauskas2016}
A.~Miliauskas and H.~Anderson.
\newblock Uncertainty framing and the {IPCC} {F}ifth {A}ssessment {R}eport.
\newblock \emph{Australian Journalism Review}, 38\penalty0 (2):\penalty0
  143--154, 2016.

\bibitem[Molina and Abadal(2021)]{Molina2021}
T.~Molina and E.~Abadal.
\newblock The evolution of communicating the uncertainty of climate change to
  policymakers: A study of {IPCC} synthesis reports.
\newblock \emph{Sustainability (Switzerland)}, 13\penalty0 (5):\penalty0 1--12,
  2021.
\newblock \doi{10.3390/su13052466}.

\bibitem[Montford(2012)]{Montford2012}
A.~W. Montford.
\newblock \emph{Hiding the Decline}.
\newblock Anglosphere Books, London, 2012.

\bibitem[Montford(2015)]{Montford2015}
A.~W. Montford.
\newblock \emph{The Hockey Stick Illusion}.
\newblock Anglosphere Books, London, 2015.

\bibitem[Moss(2011)]{Moss2011}
R.~Moss.
\newblock Reducing doubt about uncertainty: Guidance for {IPCC}'s {T}hird
  {A}ssessment.
\newblock \emph{Climatic Change}, 108\penalty0 (4):\penalty0 641--658, 2011.
\newblock \doi{10.1007/s10584-011-0182-x}.

\bibitem[Moss et~al.(2010)Moss, Edmonds, Hibbard, Manning, Rose, Van~Vuuren,
  Carter, Emori, Kainuma, Kram, Meehl, Mitchell, Nakicenovic, Riahi, Smith,
  Stouffer, Thomson, Weyant, and Wilbanks]{Moss2010}
R.~H. Moss, J.~A. Edmonds, K.~A. Hibbard, M.~R. Manning, S.~K. Rose, D.~P.
  Van~Vuuren, T.~R. Carter, S.~Emori, M.~Kainuma, T.~Kram, G.~A. Meehl,
  J.~F.~B. Mitchell, N.~Nakicenovic, K.~Riahi, S.~J. Smith, R.~J. Stouffer,
  A.~M. Thomson, J.~P. Weyant, and T.~J. Wilbanks.
\newblock The next generation of scenarios for climate change research and
  assessment.
\newblock \emph{Nature}, 463\penalty0 (7282):\penalty0 747--756, 2010.
\newblock \doi{10.1038/nature08823}.

\bibitem[Nakicenovic and Swart(2000)]{SRES}
N.~Nakicenovic and R.~Swart, editors.
\newblock \emph{Special Report on Emissions Scenarios: A special report of
  Working Group III of the Intergovernmental Panel on Climate Change}.
\newblock Cambridge University Press, Cambridge, 2000.

\bibitem[Nakicenovic et~al.(2003)Nakicenovic, Grübler, Gaffin, Jung, Kram,
  Morita, Pither, Riahi, Schlesinger, Shukla, Van~Vuuren, Davis, Michaelis,
  Swart, and Victor]{Nakicenovic2003}
N.~Nakicenovic, A.~Grübler, S.~Gaffin, T.~Jung, T.~Kram, T.~Morita, H.~Pither,
  K.~Riahi, M.~Schlesinger, P.~Shukla, D.~Van~Vuuren, G.~Davis, L.~Michaelis,
  R.~Swart, and N.~Victor.
\newblock {IPCC} {SRES} revisited: A response.
\newblock \emph{Energy and Environment}, 14\penalty0 (2-3):\penalty0 187--214,
  2003.
\newblock \doi{10.1260/095830503765184592}.

\bibitem[Narita(2012)]{Narita2012}
D.~Narita.
\newblock Managing uncertainties: The making of the {IPCC}'s {Special Report on
  Carbon Dioxide Capture and Storage}.
\newblock \emph{Public Understanding of Science}, 21\penalty0 (1):\penalty0
  84--100, 2012.
\newblock \doi{10.1177/0963662510367710}.

\bibitem[Nature(1995)]{Nature1995}
Nature.
\newblock Global warming rows.
\newblock \emph{Nature}, 378\penalty0 (6555):\penalty0 322, 1995.
\newblock \doi{10.1038/378322a0}.

\bibitem[Nature(2010)]{Nature2010}
Nature.
\newblock {IPCC}: Cherish it, tweak it or scrap it?
\newblock \emph{Nature}, 463\penalty0 (7282):\penalty0 730--732, 2010.
\newblock \doi{10.1038/463730a}.

\bibitem[{New Scientist}(2010)]{NewScientist2010}
{New Scientist}.
\newblock Let the sunlight in on climate change.
\newblock \emph{New Scientist}, 205\penalty0 (2745):\penalty0 5, 2010.
\newblock \doi{10.1016/S0262-4079(10)60196-0}.

\bibitem[Nishioka(2008)]{Nishioka2008}
S.~Nishioka.
\newblock How did science and {IPCC} lead policy?
\newblock \emph{Atomos}, 50\penalty0 (9):\penalty0 557--561, 2008.
\newblock \doi{10.3327/jaesjb.50.9_557}.

\bibitem[Nordhaus(1982)]{Nordhaus1982}
W.~D. Nordhaus.
\newblock How fast should we graze the global commons?
\newblock \emph{American Economic Review}, 72\penalty0 (2):\penalty0 242--246,
  1982.

\bibitem[Nordhaus(1992)]{Nordhaus1992}
W.~D. Nordhaus.
\newblock An optimal transition path for controlling greenhouse gases.
\newblock \emph{Science}, 258:\penalty0 1315--1319, 1992.

\bibitem[Nordhaus(2014)]{Nordhaus2014}
W.~D. Nordhaus.
\newblock {Estimates of the Social Cost of Carbon: Concepts and Results from
  the {DICE-2013R} Model and Alternative Approaches}.
\newblock \emph{Journal of the Association of Environmental and Resource
  Economists}, 1\penalty0 (1), 2014.
\newblock \doi{10.1086/676035}.

\bibitem[O'Neill and Nakicenovic(2008)]{ONeill2008a}
B.~O'Neill and N.~Nakicenovic.
\newblock Learning from global emissions scenarios.
\newblock \emph{Environmental Research Letters}, 3\penalty0 (4), 2008.
\newblock \doi{10.1088/1748-9326/3/4/045014}.

\bibitem[O'Neill et~al.(2008)O'Neill, Pulver, Vandeveer, and Garb]{ONeill2008b}
B.~O'Neill, S.~Pulver, S.~Vandeveer, and Y.~Garb.
\newblock Where next with global environmental scenarios?
\newblock \emph{Environmental Research Letters}, 3\penalty0 (4), 2008.
\newblock \doi{10.1088/1748-9326/3/4/045012}.

\bibitem[O'Neill et~al.(2014)O'Neill, Kriegler, Riahi, Ebi, Hallegatte, Carter,
  Mathur, and van Vuuren]{ONeill2014}
B.~C. O'Neill, E.~Kriegler, K.~Riahi, K.~L. Ebi, S.~Hallegatte, T.~R. Carter,
  R.~Mathur, and D.~P. van Vuuren.
\newblock A new scenario framework for climate change research: The concept of
  shared socioeconomic pathways.
\newblock \emph{Climatic Change}, 122\penalty0 (3):\penalty0 387--400, 2014.
\newblock \doi{10.1007/s10584-013-0905-2}.

\bibitem[Patt and Dessai(2005)]{Patt2005}
A.~Patt and S.~Dessai.
\newblock Communicating uncertainty: Lessons learned and suggestions for
  climate change assessment.
\newblock \emph{Comptes Rendus - Geoscience}, 337\penalty0 (4):\penalty0
  425--441, 2005.
\newblock \doi{10.1016/j.crte.2004.10.004}.

\bibitem[Patt et~al.(2022)Patt, Rajamani, Bhandari, Boncheva, Caparr\'{o}s,
  Kamal Djemouai, Kubota, Peel, Sari, F. Sprinz, and Wettestad]{Patt2022IPCC}
A.~Patt, L.~Rajamani, P.~Bhandari, A.~I. Boncheva, A.~Caparr\'{o}s,
  Kamal Djemouai, I.~Kubota, J.~Peel, A.~P. Sari, D.~F. Sprinz, and
  J.~Wettestad.
\newblock International cooperation.
\newblock In P.~R. Shukla, J.~Skea, R.~Slade, A.~A. Khourdajie, R.~van Diemen,
  D.~McCollum, M.~Pathak, S.~Some, P.~Vyas, R.~Fradera, M.~Belkacemi,
  A.~Hasija, G.~Lisboa, S.~Luz, and J.~Malley, editors, \emph{Climate Change
  2022: Mitigation of Climate Change\textemdash Contribution of Working Group
  III to the Sixth Assessment Report of the Intergovernmental Panel on Climate
  Change}. Cambridge University Press, Cambridge, 2022.

\bibitem[Pearce(1995{\natexlab{a}})]{Pearce1995c}
D.~W. Pearce.
\newblock Valuing climate change.
\newblock \emph{Chemistry and Industry}, 1995{\natexlab{a}}.

\bibitem[Pearce et~al.(1996)Pearce, Cline, Achanta, Fankhauser, Pachauri, Tol,
  and Vellinga]{Pearce1996IPCC}
D.~W. Pearce, W.~R. Cline, A.~N. Achanta, S.~Fankhauser, R.~K. Pachauri,
  R.~S.~J. Tol, and P.~Vellinga.
\newblock The social costs of climate change: Greenhouse damage and the
  benefits of control.
\newblock In J.~P. Bruce, H.~Lee, and E.~F. Haites, editors, \emph{Climate
  Change 1995: Economic and Social Dimensions\textemdash Contribution of
  Working Group III to the Second Assessment Report of the Intergovernmental
  Panel on Climate Change}, pages 179--224. Cambridge University Press,
  Cambridge, 1996.

\bibitem[Pearce(1995{\natexlab{b}})]{Pearce1995a}
F.~Pearce.
\newblock Global row over value of human life.
\newblock \emph{New Scientist}, 1991\penalty0 (AUGUST 19):\penalty0 7,
  1995{\natexlab{b}}.

\bibitem[Pearce(1995{\natexlab{c}})]{Pearce1995b}
F.~Pearce.
\newblock Price of life sends temperatures soaring.
\newblock \emph{New Scientist}, page~5, 1995{\natexlab{c}}.

\bibitem[Peiser(2007)]{Peiser2007}
B.~Peiser.
\newblock {IPCC}: The only game in town?
\newblock \emph{Energy and Environment}, 18\penalty0 (7-8):\penalty0 i--iii,
  2007.
\newblock \doi{10.1260/095830507782616850}.

\bibitem[Pereira et~al.(2022)Pereira, Ba\u{s}i\'{c}, Bogunovic, and
  Barcelo]{Pereira2022}
P.~Pereira, F.~Ba\u{s}i\'{c}, I.~Bogunovic, and D.~Barcelo.
\newblock Russian-{U}krainian war impacts the total environment.
\newblock \emph{Science of the Total Environment}, 837, 2022.
\newblock \doi{10.1016/j.scitotenv.2022.155865}.

\bibitem[{Pielke Jr} and Ritchie(2021)]{Pielke2021}
R.~A. {Pielke Jr} and J.~Ritchie.
\newblock Distorting the view of our climate future: The misuse and abuse of
  climate pathways and scenarios.
\newblock \emph{Energy Research \& Social Science}, 72:\penalty0 101890, 2021.
\newblock ISSN 2214-6296.
\newblock \doi{https://doi.org/10.1016/j.erss.2020.101890}.

\bibitem[Pielke~Jr et~al.(2008)Pielke~Jr, Wigley, and Green]{Pielke2008}
R.~A. Pielke~Jr, T.~M.~L. Wigley, and C.~Green.
\newblock Dangerous assumptions.
\newblock \emph{Nature}, 452\penalty0 (7187):\penalty0 531--532, 2008.
\newblock \doi{10.1038/452531a}.

\bibitem[{Pielke Jr} et~al.(2022){Pielke Jr}, Burgess, and Ritchie]{Pielke2022}
R.~A. {Pielke Jr}, M.~G. Burgess, and J.~Ritchie.
\newblock Plausible 2005{\textendash}2050 emissions scenarios project between 2
  {\textdegree}c and 3 {\textdegree}c of warming by 2100.
\newblock \emph{Environmental Research Letters}, 17\penalty0 (2):\penalty0
  024027, 2022.
\newblock \doi{10.1088/1748-9326/ac4ebf}.

\bibitem[Rafaty et~al.(2020)Rafaty, Dolphin, and Pretis]{Rafaty2020}
R.~Rafaty, G.~Dolphin, and F.~Pretis.
\newblock Carbon pricing and the elasticity of {CO}\textsubscript{2} emissions.
\newblock Technical report, Energy Policy Research Group, University of
  Cambridge, 2020.
\newblock URL \url{http://www.jstor.org/stable/resrep30490}.

\bibitem[Riahi et~al.(2017)Riahi, van Vuuren, Kriegler, Edmonds, O’Neill,
  Fujimori, Bauer, Calvin, Dellink, Fricko, Lutz, Popp, Cuaresma, KC, Leimbach,
  Jiang, Kram, Rao, Emmerling, Ebi, Hasegawa, Havlik, Humpenöder, Silva,
  Smith, Stehfest, Bosetti, Eom, Gernaat, Masui, Rogelj, Strefler, Drouet,
  Krey, Luderer, Harmsen, Takahashi, Baumstark, Doelman, Kainuma, Klimont,
  Marangoni, Lotze-Campen, Obersteiner, Tabeau, and Tavoni]{Riahi2017}
K.~Riahi, D.~P. van Vuuren, E.~Kriegler, J.~Edmonds, B.~C. O’Neill,
  S.~Fujimori, N.~Bauer, K.~Calvin, R.~Dellink, O.~Fricko, W.~Lutz, A.~Popp,
  J.~C. Cuaresma, S.~KC, M.~Leimbach, L.~Jiang, T.~Kram, S.~Rao, J.~Emmerling,
  K.~Ebi, T.~Hasegawa, P.~Havlik, F.~Humpenöder, L.~A.~D. Silva, S.~Smith,
  E.~Stehfest, V.~Bosetti, J.~Eom, D.~Gernaat, T.~Masui, J.~Rogelj,
  J.~Strefler, L.~Drouet, V.~Krey, G.~Luderer, M.~Harmsen, K.~Takahashi,
  L.~Baumstark, J.~C. Doelman, M.~Kainuma, Z.~Klimont, G.~Marangoni,
  H.~Lotze-Campen, M.~Obersteiner, A.~Tabeau, and M.~Tavoni.
\newblock The shared socioeconomic pathways and their energy, land use, and
  greenhouse gas emissions implications: An overview.
\newblock \emph{Global Environmental Change}, 42:\penalty0 153 -- 168, 2017.
\newblock ISSN 0959-3780.
\newblock \doi{https://doi.org/10.1016/j.gloenvcha.2016.05.009}.
\newblock URL
  \url{http://www.sciencedirect.com/science/article/pii/S0959378016300681}.

\bibitem[Riahi et~al.(2022)Riahi, Schaeffer, Arango, Calvin, Guivarch,
  Hasegawa, Jiang, Kriegler, Matthews, Peters, Rao, Robertson, Sebbit,
  Steinberger, Tavoni, and van Vuuren]{Riahi2022IPCC}
K.~Riahi, R.~Schaeffer, J.~Arango, K.~Calvin, C.~Guivarch, T.~Hasegawa,
  K.~Jiang, E.~Kriegler, R.~Matthews, G.~Peters, A.~Rao, S.~Robertson, A.~M.
  Sebbit, J.~Steinberger, M.~Tavoni, and D.~van Vuuren.
\newblock Mitigation pathways compatible with long-term goals.
\newblock In P.~R. Shukla, J.~Skea, R.~Slade, A.~A. Khourdajie, R.~van Diemen,
  D.~McCollum, M.~Pathak, S.~Some, P.~Vyas, R.~Fradera, M.~Belkacemi,
  A.~Hasija, G.~Lisboa, S.~Luz, and J.~Malley, editors, \emph{Climate Change
  2022: Mitigation of Climate Change\textemdash Contribution of Working Group
  III to the Sixth Assessment Report of the Intergovernmental Panel on Climate
  Change}. Cambridge University Press, Cambridge, 2022.

\bibitem[Richels et~al.(2008)Richels, Tol, and Yohe]{Richels2008}
R.~G. Richels, R.~S.~J. Tol, and G.~W. Yohe.
\newblock Future scenarios for emissions need continual adjustment.
\newblock \emph{Nature}, 453\penalty0 (7192):\penalty0 155, 2008.
\newblock \doi{10.1038/453155a}.

\bibitem[Risbey and Kandlikar(2007)]{Risbey2007}
J.~S. Risbey and M.~Kandlikar.
\newblock Expressions of likelihood and confidence in the {IPCC} uncertainty
  assessment process.
\newblock \emph{Climatic Change}, 85\penalty0 (1-2):\penalty0 19--31, 2007.
\newblock \doi{10.1007/s10584-007-9315-7}.

\bibitem[Ritchie and Dowlatabadi(2017)]{Ritchie2017}
J.~Ritchie and H.~Dowlatabadi.
\newblock The 1000 {GtC} coal question: Are cases of vastly expanded future
  coal combustion still plausible?
\newblock \emph{Energy Economics}, 65:\penalty0 16 -- 31, 2017.
\newblock \doi{https://doi.org/10.1016/j.eneco.2017.04.015}.

\bibitem[Rose et~al.(2022)Rose, Diaz, Carleton, Drouet, Guivarch, M\'{e}jean,
  and Piontek]{Rose2022}
S.~Rose, D.~Diaz, T.~Carleton, L.~Drouet, C.~Guivarch, A.~M\'{e}jean, and
  F.~Piontek.
\newblock Cross-working group box: Estimating global economic impacts from
  climate change.
\newblock In H.-O. Pörtner, D.~Roberts, M.~Tignor, E.~Poloczanska,
  K.~Mintenbeck, A.~Alegr\'{i}a, M.~Craig, S.~Langsdorf, S.~Löschke,
  V.~Möller, A.~Okem, and B.~Rama, editors, \emph{Climate Change 2022:
  Impacts, Adaptation, and Vulnerability. Contribution of Working Group II to
  the Sixth Assessment Report of the Intergovernmental Panel on Climate
  Change}, book section~16, pages 111--116. Cambridge University Press,
  Cambridge, 2022.

\bibitem[Rothman et~al.(2009)Rothman, van Bers, Bakkes, and
  Pahl-Wostl]{Rothman2009}
D.~Rothman, C.~van Bers, J.~Bakkes, and C.~Pahl-Wostl.
\newblock How to make global assessments more effective: lessons from the
  assessment community.
\newblock \emph{Current Opinion in Environmental Sustainability}, 1\penalty0
  (2):\penalty0 214--218, 2009.
\newblock \doi{10.1016/j.cosust.2009.09.002}.

\bibitem[Schenk and Lensink(2007)]{Schenk2007}
N.~Schenk and S.~Lensink.
\newblock Communicating uncertainty in the {IPCC}'s greenhouse gas emissions
  scenarios.
\newblock \emph{Climatic Change}, 82\penalty0 (3-4):\penalty0 293--308, 2007.
\newblock \doi{10.1007/s10584-006-9194-3}.

\bibitem[Schiermeier(2010)]{Schiermeier2010}
Q.~Schiermeier.
\newblock {IPCC} flooded by criticism.
\newblock \emph{Nature}, 463\penalty0 (7281):\penalty0 596--597, 2010.
\newblock \doi{10.1038/463596a}.

\bibitem[Sen and Vollebergh(2018)]{Sen2018}
S.~Sen and H.~Vollebergh.
\newblock {The effectiveness of taxing the carbon content of energy
  consumption}.
\newblock \emph{Journal of Environmental Economics and Management}, 92\penalty0
  (C):\penalty0 74--99, 2018.
\newblock \doi{10.1016/j.jeem.2018.08.01}.

\bibitem[Shapiro et~al.(2010)Shapiro, Diab, {de Brito Cruz}, Cropper, Fang,
  Fresco, Manabe, Mehta, Molina, Williams, Winnacker, Zakri, Linn, Elliott,
  Kearney, Leckie, Nguyen, Ortego, and Symmes]{InterAcademy2010}
H.~T. Shapiro, R.~Diab, C.~H. {de Brito Cruz}, M.~Cropper, J.~Fang, L.~O.
  Fresco, S.~Manabe, G.~Mehta, M.~Molina, P.~Williams, E.-L. Winnacker, A.~H.
  Zakri, A.~Linn, T.~Elliott, W.~Kearney, S.~Leckie, T.~Nguyen, J.~Ortego, and
  G.~Symmes.
\newblock Climate change assessments: Review of the processes and procedures of
  the {IPCC}.
\newblock Technical report, InterAcademy Council, Trieste, 2010.
\newblock URL
  \url{https://www.interacademies.org/sites/default/files/publication/climate_change_assessments_review_of_the_processes_procedures_of_the_ipcc.pdf}.

\bibitem[Shukla et~al.(2022)Shukla, Skea, Slade, Khourdajie, van Diemen,
  McCollum, Pathak, Some, Vyas, Fradera, Belkacemi, Hasija, Lisboa, Luz, and
  Malley]{IPCCWG32022}
P.~R. Shukla, J.~Skea, R.~Slade, A.~A. Khourdajie, R.~van Diemen, D.~McCollum,
  M.~Pathak, S.~Some, P.~Vyas, R.~Fradera, M.~Belkacemi, A.~Hasija, G.~Lisboa,
  S.~Luz, and J.~Malley, editors.
\newblock \emph{Climate Change 2022: Mitigation of Climate Change\textemdash
  Contribution of Working Group III to the Sixth Assessment Report of the
  Intergovernmental Panel on Climate Change}.
\newblock Cambridge University Press, Cambridge, 2022.

\bibitem[Smith(1776)]{Smith1776}
A.~Smith.
\newblock \emph{An inquiry into the nature and causes of the wealth of
  nations}.
\newblock W. Strahan and T. Cadell, London, 1776.

\bibitem[Smith et~al.(2001)Smith, Schellnhuber, Mirza, Fankhauser, Leemans,
  Erda, Ogallo, Pittock, Richels, Rosenzweig, Safriel, Tol, Weyant, and
  Yohe]{Smith2001}
J.~B. Smith, H.-J. Schellnhuber, M.~Q. Mirza, S.~Fankhauser, R.~Leemans,
  L.~Erda, L.~Ogallo, A.~B. Pittock, R.~G. Richels, C.~Rosenzweig, U.~Safriel,
  R.~S.~J. Tol, J.~P. Weyant, and G.~W. Yohe.
\newblock Vulnerability to climate change and reasons for concern: A synthesis.
\newblock In J.~J. McCarthy, O.~F. Canziani, N.~A. Leary, D.~J. Dokken, and
  K.~S. White, editors, \emph{Climate Change 2001: Impacts, Adaptation, and
  Vulnerability\textemdash Contribution of Working Group II to the Third
  Assessment Report of the Intergovernmental Panel on Climate Change}, pages
  913--967. Press Syndicate of the University of Cambridge, Cambridge, UK,
  2001.

\bibitem[Srikrishnan et~al.(2022)Srikrishnan, Guan, Tol, and
  Keller]{Srikrishnan2022}
V.~Srikrishnan, Y.~Guan, R.~S.~J. Tol, and K.~Keller.
\newblock Probabilistic projections of baseline twenty-first century
  {CO}\textsubscript{2} emissions using a simple calibrated integrated
  assessment model.
\newblock \emph{Climatic Change}, 170\penalty0 (3-4), 2022.
\newblock \doi{10.1007/s10584-021-03279-7}.

\bibitem[Stavins(2014)]{Stavins2014}
R.~N. Stavins.
\newblock Understanding the {IPCC}'s products.
\newblock \emph{Environmental Forum}, 31\penalty0 (3):\penalty0 14, 2014.

\bibitem[Stavins(2015)]{Stavins2015}
R.~N. Stavins.
\newblock The {IPCC} at a crossroads.
\newblock \emph{Environmental Forum}, 32\penalty0 (3):\penalty0 16, 2015.

\bibitem[Swart et~al.(2009)Swart, Bernstein, Ha-Duong, and Petersen]{Swart2009}
R.~Swart, L.~Bernstein, M.~Ha-Duong, and A.~Petersen.
\newblock Agreeing to disagree: Uncertainty management in assessing climate
  change, impacts and responses by the {IPCC}.
\newblock \emph{Climatic Change}, 92\penalty0 (1-2):\penalty0 1--29, 2009.
\newblock \doi{10.1007/s10584-008-9444-7}.

\bibitem[Tavoni and Tol(2010)]{Tavoni2010}
M.~Tavoni and R.~S.~J. Tol.
\newblock Counting only the hits? the risk of underestimating the costs of
  stringent climate policy: A letter.
\newblock \emph{Climatic Change}, 100\penalty0 (3):\penalty0 769--778, 2010.
\newblock \doi{10.1007/s10584-010-9867-9}.

\bibitem[Tirpak and Vellinga(1990)]{Tirpak1990}
D.~Tirpak and P.~Vellinga.
\newblock Emissions scenarios.
\newblock In F.~Bernthal, E.~Dowdeswell, J.~Luo, D.~Attard, P.~Vellinga, and
  R.~Karimanzira, editors, \emph{Climate Change: The {IPCC} Response
  Strategies}, chapter~2, pages 9--44. Cambridge University Press, Cambridge,
  1990.

\bibitem[Tol(1997)]{Tol1997}
R.~S.~J. Tol.
\newblock On the optimal control of carbon dioxide emissions\textemdash an
  application of {FUND}.
\newblock \emph{Environmental Modelling and Assessment}, 2:\penalty0 151--163,
  1997.

\bibitem[Tol(1999)]{Tol1999kyoto}
R.~S.~J. Tol.
\newblock Kyoto, efficiency, and cost-effectiveness: Applications of {FUND}.
\newblock \emph{The Energy Journal}, \penalty0 (Special Issue):\penalty0
  131--156, 1999.
\newblock URL \url{https://ideas.repec.org/a/aen/journl/1999si-a06.html}.

\bibitem[Tol(2006)]{Tol2006}
R.~S.~J. Tol.
\newblock Exchange rates and climate change: An application of {FUND}.
\newblock \emph{Climatic Change}, 75\penalty0 (1-2):\penalty0 59--80, 2006.
\newblock \doi{10.1007/s10584-005-9003-4}.

\bibitem[Tol(2011)]{Tol2011CC}
R.~S.~J. Tol.
\newblock Regulating knowledge monopolies: The case of the {IPCC}.
\newblock \emph{Climatic Change}, 108\penalty0 (4):\penalty0 827--839, 2011.
\newblock \doi{10.1007/s10584-011-0214-6}.

\bibitem[Tol(2012)]{Tol2012EP}
R.~S.~J. Tol.
\newblock A cost-benefit analysis of the {EU} 20/20/2020 package.
\newblock \emph{Energy Policy}, 49:\penalty0 288--295, 2012.
\newblock \doi{10.1016/j.enpol.2012.06.018}.

\bibitem[Tol(2013)]{Tol2013China}
R.~S.~J. Tol.
\newblock Low probability, high impact: The implications of a break-up of
  {C}hina for carbon dioxide emissions.
\newblock \emph{Climatic Change}, 117\penalty0 (4):\penalty0 961--970, 2013.
\newblock \doi{10.1007/s10584-013-0723-6}.

\bibitem[Tol(2014)]{Tol2014En}
R.~S.~J. Tol.
\newblock Ambiguity reduction by objective model selection, with an application
  to the costs of the {EU} 2030 climate targets.
\newblock \emph{Energies}, 7\penalty0 (11):\penalty0 6886--6896, 2014.
\newblock URL \url{https://www.mdpi.com/1996-1073/7/11/6886}.

\bibitem[Tol(2016)]{Tol2016}
R.~S.~J. Tol.
\newblock The impacts of climate change according to the {IPCC}.
\newblock \emph{Climate Change Economics}, 7\penalty0 (1), 2016.
\newblock \doi{10.1142/S2010007816400042}.

\bibitem[Tol(2018)]{Tol2018REEP}
R.~S.~J. Tol.
\newblock The economic impacts of climate change.
\newblock \emph{Review of Environmental Economics and Policy}, 12\penalty0
  (1):\penalty0 4--25, 2018.
\newblock URL \url{http://dx.doi.org/10.1093/reep/rex027}.

\bibitem[Tol(2019)]{Tol2019bk}
R.~S.~J. Tol.
\newblock \emph{Climate economics: Economic analyses of climate, climate
  change, and climate policy (second edition)}.
\newblock Edward Elgar, Cheltenham, 2019.

\bibitem[Tol(2020)]{Tol2020CCE}
R.~S.~J. Tol.
\newblock Selfish bureaucrats and policy heterogeneity in {N}ordhaus' {DICE}.
\newblock \emph{Climate Change Economics}, 2020.
\newblock \doi{10.1142/S2010007820400060}.

\bibitem[Tol(2021)]{Tol2021IE}
R.~S.~J. Tol.
\newblock Europe’s climate target for 2050: An assessment.
\newblock \emph{Intereconomics}, 56\penalty0 (6):\penalty0 330--335, 2021.
\newblock \doi{10.1007/s10272-021-1012-7}.

\bibitem[Tol et~al.(2012)Tol, Berntsen, Oneill, Fuglestvedt, and
  Shine]{Tol2012ERL}
R.~S.~J. Tol, T.~K. Berntsen, B.~C. Oneill, J.~S. Fuglestvedt, and K.~P. Shine.
\newblock A unifying framework for metrics for aggregating the climate effect
  of different emissions.
\newblock \emph{Environmental Research Letters}, 7\penalty0 (4), 2012.
\newblock \doi{10.1088/1748-9326/7/4/044006}.

\bibitem[{UNFCCC}(2021)]{UNFCCC2021}
{UNFCCC}.
\newblock {NDC} {S}ynthesis {R}eport.
\newblock Technical report, United Nations Framework Convention on Climate
  Change Secretariat, 2021.
\newblock URL
  \url{https://unfccc.int/process-and-meetings/the-paris-agreement/nationally-determined-contributions-ndcs/nationally-determined-contributions-ndcs/ndc-synthesis-report}.

\bibitem[van~der Sluijs(2012)]{Sluijs2012}
J.~P. van~der Sluijs.
\newblock Uncertainty and dissent in climate risk assessment: A post-normal
  perspective.
\newblock \emph{Nature and Culture}, 7\penalty0 (2):\penalty0 174--195, 2012.
\newblock \doi{10.3167/nc.2012.070204}.

\bibitem[Van~Vuuren and Alfsen(2006)]{VanVuuren2006}
D.~Van~Vuuren and K.~Alfsen.
\newblock {PPP} versus {MER}: Searching for answers in a multi-dimensional
  debate.
\newblock \emph{Climatic Change}, 75\penalty0 (1-2):\penalty0 47--57, 2006.
\newblock \doi{10.1007/s10584-005-9045-7}.

\bibitem[Wegman et~al.(2010)Wegman, Scott, and Said]{Wegman2010}
E.~J. Wegman, D.~W. Scott, and Y.~H. Said.
\newblock Ad hoc committee report on the ‘hockey stick’ global climate
  reconstruction.
\newblock Technical report, Science \& Public Policy Institute, 2010.
\newblock URL
  \url{https://www.science-e-publishing.de/project/lv-twk/images/pdfs/2010_Wegman_ad_hoc_report.pdf}.

\bibitem[Weyant et~al.(1996)Weyant, 0gunlade Davidson, Dowlatabadi, Edmonds,
  Grubb, Parson, Richels, Rotmans, Shukla, Tol, Cline, and
  Fankhauser]{Weyant1996IPCC}
J.~P. Weyant, 0gunlade Davidson, H.~Dowlatabadi, J.~A. Edmonds, M.~L. Grubb,
  E.~A. Parson, R.~G. Richels, J.~Rotmans, P.~R. Shukla, R.~S.~J. Tol, W.~R.
  Cline, and S.~Fankhauser.
\newblock Integrated assessment of climate change: An overview and comparison
  of approaches and results.
\newblock In J.~P. Bruce, H.~Lee, and E.~F. Haites, editors, \emph{Climate
  Change 1995: Economic and Social Dimensions of Climate Change\textemdash
  Contribution of Working Group III to the Second Assessment Report of the
  Intergovernmental Panel on Climate Change}, chapter~4. Cambridge University
  Press, Cambridge, 1996.

\bibitem[{WMO}(2022)]{WMO2022}
{WMO}.
\newblock Global annual to decadal climate update.
\newblock Technical report, World Meteorological Organization, Geneva, 2022.
\newblock URL
  \url{https://hadleyserver.metoffice.gov.uk/wmolc/WMO_GADCU_2022-2026.pdf}.

\bibitem[Xing et~al.(2021)Xing, Leard, and Li]{XING2021}
J.~Xing, B.~Leard, and S.~Li.
\newblock What does an electric vehicle replace?
\newblock \emph{Journal of Environmental Economics and Management},
  107:\penalty0 102432, 2021.
\newblock \doi{https://doi.org/10.1016/j.jeem.2021.102432}.

\bibitem[Yuan et~al.(2022)Yuan, Rodrigues, Tukker, and Behrens]{Yuan2022}
R.~Yuan, J.~F. Rodrigues, A.~Tukker, and P.~Behrens.
\newblock The statistical projection of global ghg emissions from a consumption
  perspective.
\newblock \emph{Sustainable Production and Consumption}, 2022.
\newblock ISSN 2352-5509.
\newblock \doi{https://doi.org/10.1016/j.spc.2022.09.021}.
\newblock URL
  \url{https://www.sciencedirect.com/science/article/pii/S2352550922002615}.

\bibitem[Zorita(2019)]{Zorita2019}
E.~Zorita.
\newblock The climate of the past millennium and online public engagement in a
  scientific debate.
\newblock \emph{Wiley Interdisciplinary Reviews: Climate Change}, 10\penalty0
  (5), 2019.
\newblock \doi{10.1002/wcc.590}.

\end{thebibliography}

\end{document}